# Fano Resonance of a Magnonic Ring Resonator with a Nano-Scatterer Embedded in a Magnonic Conduit

Rajesh Kumar[1,*], Yat-Yin Au[1], and Snorri Ingvarsson[1]

[1]Science Institute, University of Iceland, Dunhaga 3, IS-107 Reykjavik, Iceland

[*]rak43@hi.is

**Abstract**

We present micromagnetic simulation of a nanoscale magnonic ring resonator coupled to a magnonic waveguide, with and without a hole embedded in the waveguide (H-NRR and NRR respectively), for realizing transformation into a Fano resonance frequency spectral lineshape from a Lorentzian one. Transmission and reflection spectra of the H-NRR and NRR have been analyzed for different structural parameters (embedded hole diameter and gap between the ring and the waveguide) and material parameters (damping factor of the ring) to achieve maximum spectral transmission magnitude variation. The obtained micromagnetic simulation results are supported by an analytical model that consists of two-tracks of wave propagation, forwards and backwards. Our findings show that H-NRR system demonstrates a monotonic dependence of spectral transmission magnitude variation against the aforementioned design parameters (gap and damping) which should provide better engineering controllability and tolerance when compared with the original NRR system. Overall, our proposed structure and study may help in building unconventional magnonic integrated circuits for sensing, data processing systems and computing.

Among all the magnonic resonators[1,2,3,4,5,6,7,8,9,10], the magnonic nano-ring resonator (NRR) is one of the most essential fundamental units in on-chip integrated spintronic technology[1],[11] with a wide range of applications such as in phase shifters[1], logic gates[11] as well as in reservoir[12] and neuromorphic computing[1]. The performance of ring resonator-based devices—such as power consumption, sensitivity, and modulation depth—strongly depends on the resonance lineshape[13,14,15,16]. Fano resonances exhibit a sharp resonant slope, large transmission variation, and high tunability[17,18,19], which consequently enhances overall device performance. The Fano resonance is characterized by an asymmetric line shape arising from the interference between a discrete state and a continuum state. Among Fano-type *photonic* resonators[19,20,21], recently, Gu et al.[22] and Wang et al.[23] reported that Fano resonance–based devices using a photonic micro-ring resonator (having spectrally discrete resonant modes) coupled with a hole-embedded straight waveguide (having a broadband scatterer) offer several advantages, include simple design, large fabrication tolerance, ease of fabrication, and the ability to achieve Fano resonances at all modes, enabling broadband operational capability. However, the magnonics realization of Fano resonance using a NRR (consists of a ring that is side coupled to a bus magnonic waveguide) remains largely unexplored, but provides device miniaturization down to nanoscale, and non-linearity. Fano resonances can be realized by embedding a hole in the magnonic waveguide side-coupled to the magnonic ring. However, introducing a hole significantly enhances backward spin-wave propagation due to backscattering in the waveguide. Reflection analysis is therefore essential, as backward-propagating spin waves (SWs) can significantly affect the performance of other devices in the same magnonic integrated circuit. Moreover, previous studies on NRRs have primarily considered only the forward propagation of SWs while neglecting the backward propagation induced by the ring structure[1],[11].

Here, we have demonstrated the design and operation of the NRR and a hole-embedded NRR (H-NRR), utilizing micromagnetic simulations. For the NRR and H-NRR structures, SW profile, transmission and reflection spectra are systematically analyzed to understand the behavior of the resonance line shape. An analytical model is developed to describe the frequency spectra characteristics of the NRR and H-NRR in the linear regime, and the results are validated through comprehensive micromagnetic simulations. Furthermore, the H-NRR has been discussed from the perspective of device fabrication and performance.

A H-NRR consists of a ring, side-coupled to a hole-embedded bus waveguide. Yttrium iron garnet (YIG) is chosen for both the ring and the waveguide due to its low damping factor, which enables long-range spin-wave propagation. The outer diameter of the ring is denoted by $D$, while the widths of both the waveguide and the ring are identical, $w = 100$ nm. The length of the waveguide is $l = 4000$ nm, and the thickness of both the waveguide and the ring is $h = 30$ nm. The gap between the ring and the waveguide is $g = 20$ nm. In this configuration, the waveguide is placed adjacent to the ring in the $x$–$y$ plane, with the input and output ports located at opposite ends of the waveguide (Figure *1*). To create the hole in bus waveguide of the H-NRR, a cylindrical hole with a nominal diameter $d_h$ is introduced at the center of the waveguide, extending through the entire thickness of the structure. In the micromagnetic simulation program MUMAX3[24], the boundary of the hole is represented by discrete grid cells, resulting in a grid-discretized circular boundary (inset of Figure *1*a). A detailed study of the shape and size of the embedded hole is presented in Section-A of the Supplementary Material (SM). The in-plane magnetization distributions of the H-NRR are shown in Figure *1*b. The details of the micromagnetic simulation have been discussed in the SM (section-B).

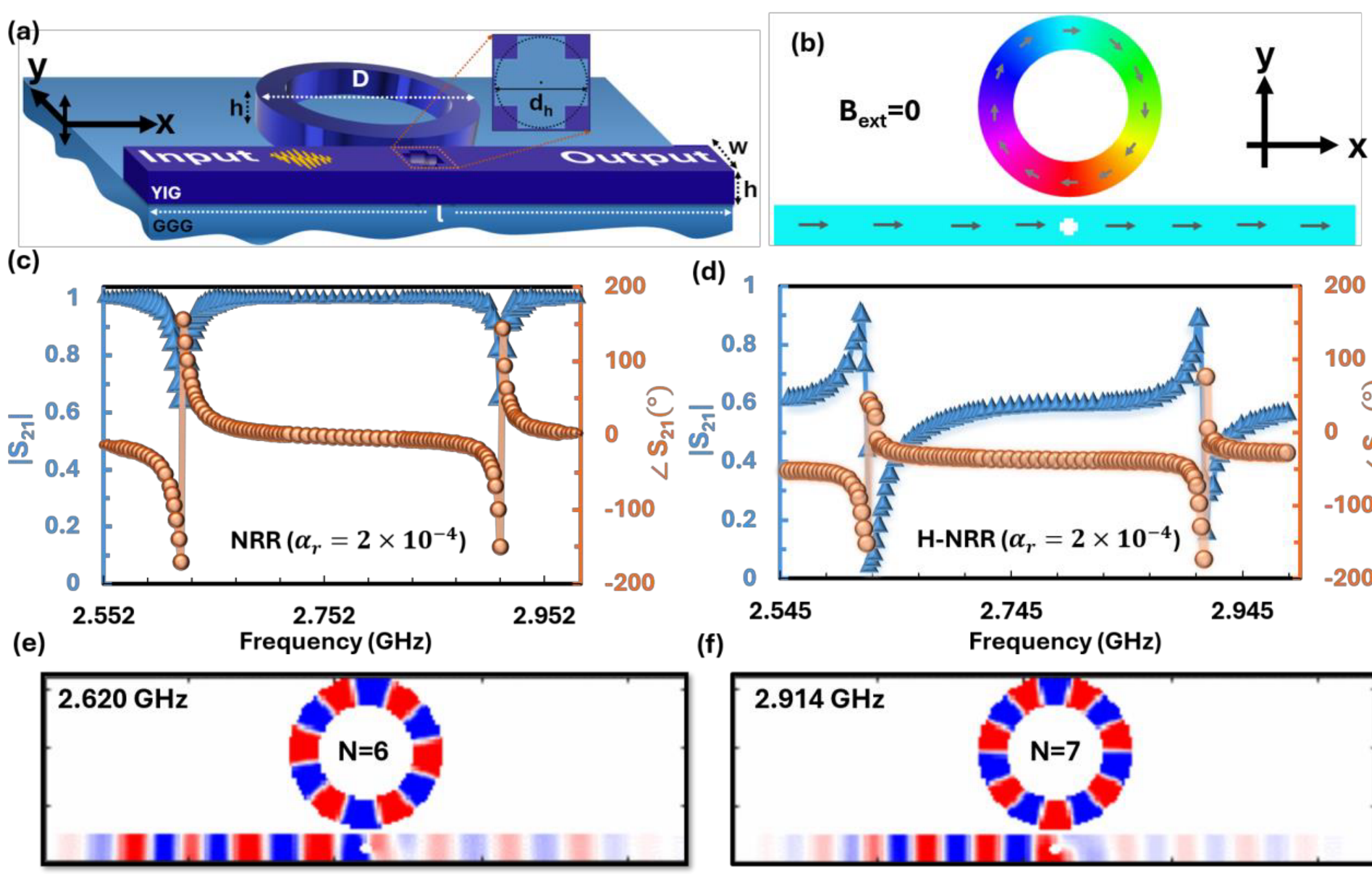

Figure 1 (a) Three-dimensional schematic illustrations of the H-NRR with geometrical parameters: $l$, $w$, $D$, $h$ and $d_h$. The insets in (a) show the top views of the embedded hole in the waveguide. (b) The initial in-plane magnetization distributions (under zero bias field) are shown for H-NRR where the gray arrows and color scale indicate the direction of the in-plane magnetization. Simulated transmission magnitude and phase spectra of (c) NRR (no hole) and (d) H-NRR computed by cw excitation method. SW profile of H-NRR at frequency (e) 2.620 GHz (transmission dip of N=6th mode) and (f) 2.914 GHz (transmission dip of N=7th mode).

To investigate the transmission and reflection spectra of the H-NRR (D = 520nm, $\alpha_r = 2 \times 10^{-4}$, $d_h = 40nm$), simulations were performed over 2.55–2.98 GHz, where Fano resonance line shapes appear at two distinct frequencies corresponding to the (N=)6th and 7th modes of the ring (Figure 1d). These resonant modes occur at frequencies ($\Omega_0$) when the corresponding $k$ number according to the dispersion relation $\omega = \omega(k)$ in the w=100 nm waveguide (see SM Section-B) satisfies the condition $kD_{sw} = 2N$, $D_{sw}$ is the diameter of the circular trajectory of SW travelling inside the ring, which naively should equal to the average between inner and outer diameter of the ring. Figure *1*e and f vindicate the above assertions by showing the SW profile of H-NRR (out-of-plane component of the spin-wave amplitude, $m_z$) at f = 2.620 GHz (transmission dip of the N=6$^{\text{th}}$ mode) and f = 2.914 GHz (transmission dip of the N=7$^{\text{th}}$ mode), respectively.

In complete of absence of the embedded hole (i.e. NRR only, or $d_h = 0$ nm), a Lorentzian lineshape is achieved (Figure *1*c). However, by introducing the hole ($d_h = 40\,nm$) in the center of the bus waveguide, the Lorentzian lineshape gets converted into a Fano lineshape (Figure *1*d). To intuitively explain this conversion of lineshape from Lorentzian to Fano, one can consider the transmission coefficient of a single resonator coupled to a waveguide[3]

$$\tau = 1 - \frac{2i\Gamma_F}{\delta\omega + i\Gamma_{\text{tot}}} \tag{1}$$

where $\delta\omega$ is the deviation of incident spin-wave frequency from the ring resonance, while $\Gamma_F$ and $\Gamma_{\text{tot}}$ are the partial and total radiative linewidth of the ring resonance. As $\delta\omega$ goes from $-\infty$ to $+\infty$, equation (1) carves out a circular trajectory in the complex plane of $\tau$ (Figure *2*a, left graph), where the start and end point is located at where $|\tau|$ is at the maximum within the trajectory. As explained in Ref. 3, $S_{21} = \tau^*$ (the asterisk denotes complex conjugation) (i.e. $|S_{21}| = |\tau|$). Therefore, a symmetric Lorentzian line shape is produced in the spectrum of $|S_{21}|$ (Figure *2*a, right graph). Now the effect of introducing a hole in the waveguide is to simply replace the "1" in the right side of equation (1) with the transmission coefficient of a hole $\tau_s$, which in general is a complex number, i.e.

$$\tau = \tau_s - \frac{2i\Gamma_F}{\delta\omega + i\Gamma_{\text{tot}}} \tag{2}$$

This results in the shift of the start and end point of the circular trajectory in the complex plane of $\tau$ from $1 + 0i$ to $\tau_s$ (Figure *2*b, left graph). Since the start and end point now is no longer located at the $|\tau|$ maximum of the trajectory, $|\tau|$ increase initially before it falls again as $\delta\omega$ increase from $-\infty$. This results in a zigzag (asymmetric) trajectory of $|S_{21}|$ against $\delta\omega$ (Figure *2*b, right graph) which is as known as the asymmetric Fano lineshape.

The single resonator model behind equation (1) and (2) unfortunately does not take into account that there exist two degenerate modes at each ring resonance, namely for clockwise and counterclockwise propagation within the ring. These two degenerate modes often hybridize with each other through their interaction with the waveguide and this hybridization is often enhanced as a result of introducing a hole into the waveguide. In order to tackle these aspects in a more comprehensive manner, an analytical toy model is developed as illustrated in Figure *2*c. The magnonic waveguide is represented by two tracks for wavefunctions $\psi_{R(L)}(x,t)$ (R and L stands for right and left going wave respectively) while the two degenerate resonance modes (propagating counterclockwise and clockwise) are simplified as two state variables $\phi_{R(L)}(t)$ with no spatial dependence. These wavefunctions and state variables obey the following equations of motion,

$$\frac{\partial\psi_{R(L)}}{\partial t} \pm v\frac{\partial\psi_{R(L)}}{\partial x} + i\Omega_k\psi_{R(L)} = -i\eta_0\Delta_F(x)\phi_{R(L)}(t) - i\eta_0\Delta_B(x)\phi_{L(R)}(t) \tag{3}$$

$$\frac{\partial\phi_{R(L)}(t)}{\partial t} + i(\Omega_0 - i\Gamma_0)\phi_{R(L)}(t) = -i\int_{-\eta/2}^{\eta/2}\Delta_F^*(x)\psi_{R(L)}(x,t)dx - i\int_{-\eta/2}^{\eta/2}\Delta_B^*(x)\psi_{L(R)}(x,t)dx \tag{4}$$

Where in equation (3) $\pm$ sign take $+(-)$ value for R (L) labelled variables respectively. At the same time, the interaction function $\Delta_{F(B)}(x)$ is given as

$$\Delta_{F(B)}(x) = \begin{cases} (1/\eta)\Delta_{F(B)} & \text{for } |x| < \eta/2 \\ 0 & \text{otherwise} \end{cases} \tag{5}$$

On the other hand, the nano-scatterer embedded in the waveguide with transmission and reflection coefficient denoted as $\tau_s$ and $r_s$ is being represented as a boundary condition at $x = 0$ given by

$$\begin{cases} \psi_R(0^+) = \tau_s\psi_R(0^-) + r_s\psi_L(0^+) \\ \psi_L(0^-) = \tau_s\psi_L(0^+) + r_s\psi_R(0^-) \end{cases} \tag{6}$$

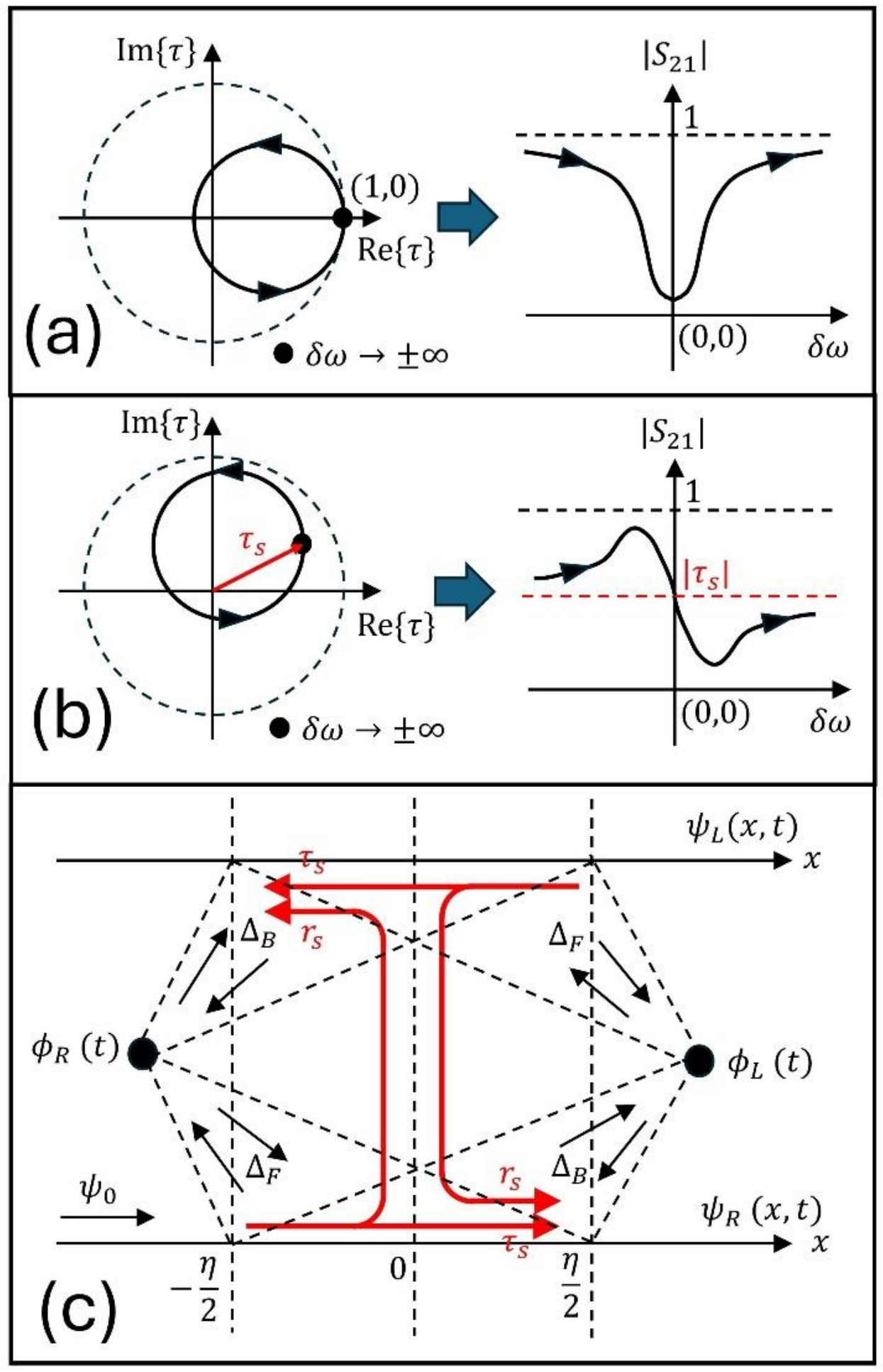

Figure 2 (a) Trajectory of $\tau$ in the complex plane according to equation (1) (left graph) that results in a symmetric Lorentzian lineshape of $|S_{21}|$ vs $\delta\omega$ (right graph). The dashed circle in the left graph represents the $|\tau| = 1$ boundary. (b) Trajectory of $\tau$ in the complex plane according to equation (2) that results in a asymmetric Fano lineshape of $|S_{21}|$ vs $\delta\omega$. (c) Illustration of the two track model with a nano-scatterer located at $x = 0$.

We write the solution to equation (3) as

$$\psi_R(x,t) = \begin{cases} \psi_0 e^{ikx} e^{-i\omega t} & \text{for } x < -\eta/2 \\ \tau\psi_0 e^{ikx} e^{-i\omega t} & \text{for } x > \eta/2 \end{cases} \tag{7}$$

$$\psi_L(x,t) = \begin{cases} r\psi_0 e^{-ikx} e^{-i\omega t} & \text{for } x < -\eta/2 \\ 0 & \text{for } x > \eta/2 \end{cases} \tag{8}$$

Then $r$ and $\tau$ can be solved for as a function of_$\omega$ and $k$,_where $\omega$ and $k$ are related by a dispersion relation determined in a separated simulation (see SM section B). The solutions for $r$ and $\tau$ as functions of $\omega$ with fit parameters $\Omega_0$, $\Gamma_0$, $\Gamma_F$, $\Gamma_B$, $\eta$, $r_S$ and $\tau_S$ and their derivation are presented in SM (section D), where we have defined $\Gamma_{F(B)} = \eta_0 |\Delta_{F(B)}|^2/(2v)$. The transmission and reflection coefficient in the simulation are then given as $S_{21} = \tau^*$ and $S_{11} = r^*$.

Micromagnetic simulation yields the transmission and reflection spectra of the NRR (with $D =$ 520 nm and $\alpha_r = 2 \times 10^{-4}$) exhibiting a resonance frequency at 2.620 GHz, as shown in Figure 3a and e (magnitude and phase as blue circles and brown triangles), respectively, in form of a symmetric Lorentzian line shape, which corresponds to the N=6th resonant mode of the structure. In the same figures, the fittings from the analytical model for the magnitude and phase of these scattering coefficients are shown as blue and brown solid curves respectively. $D_{sw}$ is the diameter of the circular trajectory of SW travelling inside the ring, which in principle should be equal to the average between inner and outer diameter of the ring (i.e., $D_{sw} = (520 + 320)/2 = 420$ nm). However, the $D_{sw}$ value required to achieve agreement of resonance frequency with the simulation data is around 470 nm (Table 1). This larger value of fitted $D_{sw}$ reflects that the SW tends to travel nearer to the outer edge of the ring, which is supported by the mode profiles of $|m_z|$ in Figure 3i. Figure 3i and j show the magnitude and phase of the spin-wave profiles (out-of-plane component of the spin-wave amplitude, $m_z$) of the NRR at frequency ($f = 2.623$ GHz), respectively, for $\alpha_r = 2 \times 10^{-3}$. The SW forms a standing wave within the ring manifested as a zebra pattern of the $|m_z|$ (Figure 3i). This indicates excitation of both the clockwise and counterclockwise propagating SW modes in the ring. For the $N$=6 mode, reflection in the waveguide is significant (zebra pattern of $|m_z|$ in the left part of the straight waveguide in Figure 3i). This is concomitant with the relatively large peak in the reflection spectrum of the NRR (Figure 3a). Conversely, for the N=7 resonant mode, the reflection is minimal (Figure 3c). This is also manifested in the weaker amplitude and relatively uniform pattern of $|m_z|$ in the left part of the waveguide region in Figure 3k.

| | Figure 3 a and e (NRR) | Figure 3 b and f (H-NRR) | Figure 3 c and g (NRR) | Figure 3 d and h (H-NRR) |
|---|---|---|---|---|
| N | 6 | 6 | 7 | 7 |
| $\eta$ | 163 nm | 110 nm | 157 nm | 110 nm |
| $D_{sw}$ | 471.0 nm | 467.0 nm | 463.6 nm | 460.5nm |
| $\Omega_0/(2\pi)$ | 2.6016 GHz | 2.6134 GHz | 2.8935 GHz | 2.9074 GHz |
| $\Gamma_0$ | $2 \times 10^{-4}\Omega_0$ | $2 \times 10^{-4}\Omega_0$ | $2 \times 10^{-4}\Omega_0$ | $2 \times 10^{-4}\Omega_0$ |
| $\Gamma_F$ | $1.408 \times 10^{-2}\Omega_0$ | $3.072 \times 10^{-3}\Omega_0$ | $1.331 \times 10^{-2}\Omega_0$ | $1.412 \times 10^{-3}\Omega_0$ |
| $\Gamma_B$ | $5.633 \times 10^{-5}\Omega_0$ | $4.095 \times 10^{-6}\Omega_0$ | $3.472 \times 10^{-7}\Omega_0$ | $1.412 \times 10^{-5}\Omega_0$ |
| $\mathrm{sign}(\sqrt{\Gamma_R\Gamma_L})$ | +1 | +1 | +1 | +1 |
| $\lambda_0$ | 246.6 nm | 244.5 nm | 208.1 nm | 206.7 nm |
| $\Gamma_F'$ | $2.5\times10^{-3}\Omega_0$ | $1.5\times10^{-3}\Omega_0$ | $1.15\times10^{-3}\Omega_0$ | $5\times10^{-4}\Omega_0$ |
| $\Gamma_B'$ | $1.0\times10^{-5}\Omega_0$ | $2.0\times10^{-6}\Omega_0$ | $3\times10^{-8}\Omega_0$ | $5\times10^{-6}\Omega_0$ |
| $\lvert r_s\rvert$ | 0 | 0.83 | 0 | 0.81 |
| $\angle r_s$ | $0^o$ | $-45^o$ | $0^o$ | $-60^o$ |
| $\lvert\tau_s\rvert$ | 1 | 0.55776 | 1 | 0.58643 |
| $\angle\tau_s$ | $0^o$ | $45^o$ | $0^o$ | $40^o$ |

Table 1 Fitting parameters for analytical model predictions in Figure 3a to h. The effective radiative linewidth $\Gamma_{F(B)}^{'}$is defined as $\Gamma_{F(B)}^{'}=\zeta^{''}\Gamma_{F(B)}$, where $\zeta^{''}=\sin^2(k\eta/2)/(k\eta/2)^2$ could be seen as the real coupling strength between SW mode in the ring with the waveguide (see SM section D).

In the H-NRR, it is worth mentioning that the effective radiative linewidth of the ring into the waveguide, i.e. $\Gamma_F'$ and $\Gamma_B'$, for N=6 has to be decreased significantly (from $\Gamma_F'=2.5\times10^{-3}$ to $1.5\times10^{-3}$ and $\Gamma_B'=1.0\times10^{-5}$ to $2.0\times10^{-6}$, see Table 1), which indicates the existence of the hole has significantly weakened the coupling of SWs between the ring and the waveguide. The interaction length $\eta$ also has to be significantly reduced (from $\eta$ =163 nm to 110 nm, see Table 1), which is reasonable as the existence of the hole has removed a large section of the waveguide that is essential to coupling to the ring.

The transmission and reflection spectra for the N=7 resonance in the ring are displayed in Figure 3 (lower enclosed region) and the corresponding fitting parameters are presented in Table 1. When comparing with the N=6 case, $\Gamma_F'$ is significantly smaller for N=7 ( $2.5\times10^{-3}\Omega_0$ vs. $1.15\times10^{-3}\Omega_0$). At the same time, $\Gamma_B'$ is also reduced from $1\times10^{-5}\Omega_0$ to$3\times10^{-8}\Omega_0$ from N=6 to N=7 resonance (see Table 1), which is a reduction of two order of magnitudes. This reduces the peak reflection magnitude from ~0.62 (Figure 3a) to less than 0.1 (Figure 3c). Such reduction is related to the wavevector matching of coupling stray field from the ring with SWs in the waveguide. Unfortunately, the two track model adopted in this article is purely phenomenological in this aspect and does not give a deeper explanation regarding this reduction of reflection strength. Although reflection is greatly reduced in the N=7 mode compared to the N=6, the minimum to maximum transition of the N=7 Fano Resonance reflection magnitude is only slightly reduced when compared to the case of N=6 (Figure 3b vs Figure 3d). The introduction of a hole in the waveguide produces significant reflection in the waveguide which enhances coupling to the clockwise mode in the ring ($\phi_L$) from $\Gamma_B'=3\times10^{-8}$ to $5\times10^{-6}$ (see Table 1).

The lack of reflection in the N=7 mode in the NRR case is manifested in the uniformity of $|m_z|$ in the left part of the waveguide in the SW mode profile (Figure 3k), and in the lack of zebra pattern inside the ring, indicating that only the counterclockwise N=7 but not the clockwise mode is being excited.

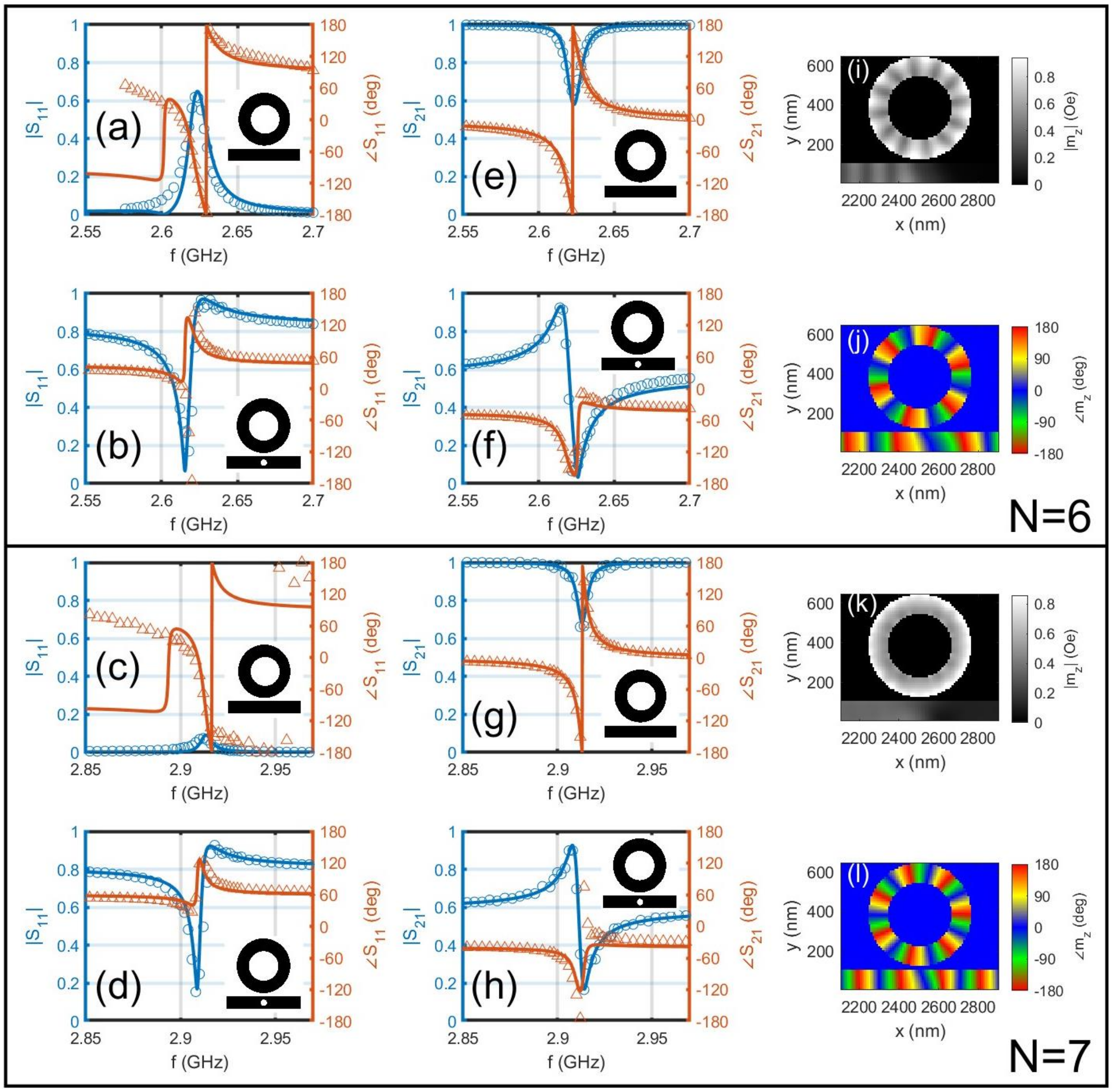

Figure 3 Simulation results of reflection ($S_{11}$) and transmission ($S_{21}$) responses showing magnitude (blue circles) and phase (brown triangles) computed by cw excitation method. Panels (a)–(d) present $S_{11}$ for NRR (no hole) and H-NRR ( $d_h = 40$ nm) at $\alpha_r = 2 \times 10^{-4}$ for modes $N = 6$ and $N = 7$, respectively. Panels (e)–(h) show the corresponding $S_{21}$ results under the same conditions. Analytical fitting of magnitude (phase) is indicated by blue (brown) solid curves, with fitting parameters listed in Table 1. Spatially pointwise temporal Fourier transform of z components of the time-dependent magnetization: (i) Magnitude, and, (j) Phase of $\mathcal{F}\{m_z(x, y, z, t)\}(f)$ at f = 2.623 GHz (N=6 resonance) for z=15 nm (middle of the YIG layer) computed using the broadband pulse excitation method for $\alpha_r = 2 \times 10^{-3}$. (k) and (l) are same as (i) and (j) except at different frequency f=2.912 GHz (N=7 resonance).

In the illustration between Figure 4(a) and (b), we define the transmission variation ($|S_{21}|_{max}$-$|S_{21}|_{min}$) for the case of NRR (Lorentzian lineshape) and the H-NRR (Fano lineshape). This transmission variation is plotted against increasing gap g value for the case of NRR and H-NRR (Figure 4a) for a fixed $\alpha_r = 2 \times 10^{-4}$. The NRR transmission variation shows an “increase follow by decrease” trend which peak at g=40 nm, which is the well-known critical coupling phenomenon, while for the case of H-NRR, transmission variation decreases monotonically with increasing g. We could also switch the place of g and $\alpha_r$ by keeping g constant at 20 nm and increase $\alpha_r$[25,26,27] in which the resultant transmission variation against $\alpha_r$ is shown in Figure 4b. Again, the NRR transmission variation against $\alpha_r$ shows a "increase-then-decrease" trend which peaks at a certain value (in this case $\alpha_r = 2 \times 10^{-3}$), while for the case of H-NRR, transmission variation decreases monotonically with increasing $\alpha_r$. Therefore, from the data of Figure 4(a) and (b), we assert that by drilling a hole in the waveguide which convert the system from NRR to H-NRR, we transform our engineering problem from searching for an optimal value of g or $\alpha_r$ somewhere in the middle of a possible range of values into getting g or $\alpha_r$as small as possible. We believe the latter task is easier to manage.

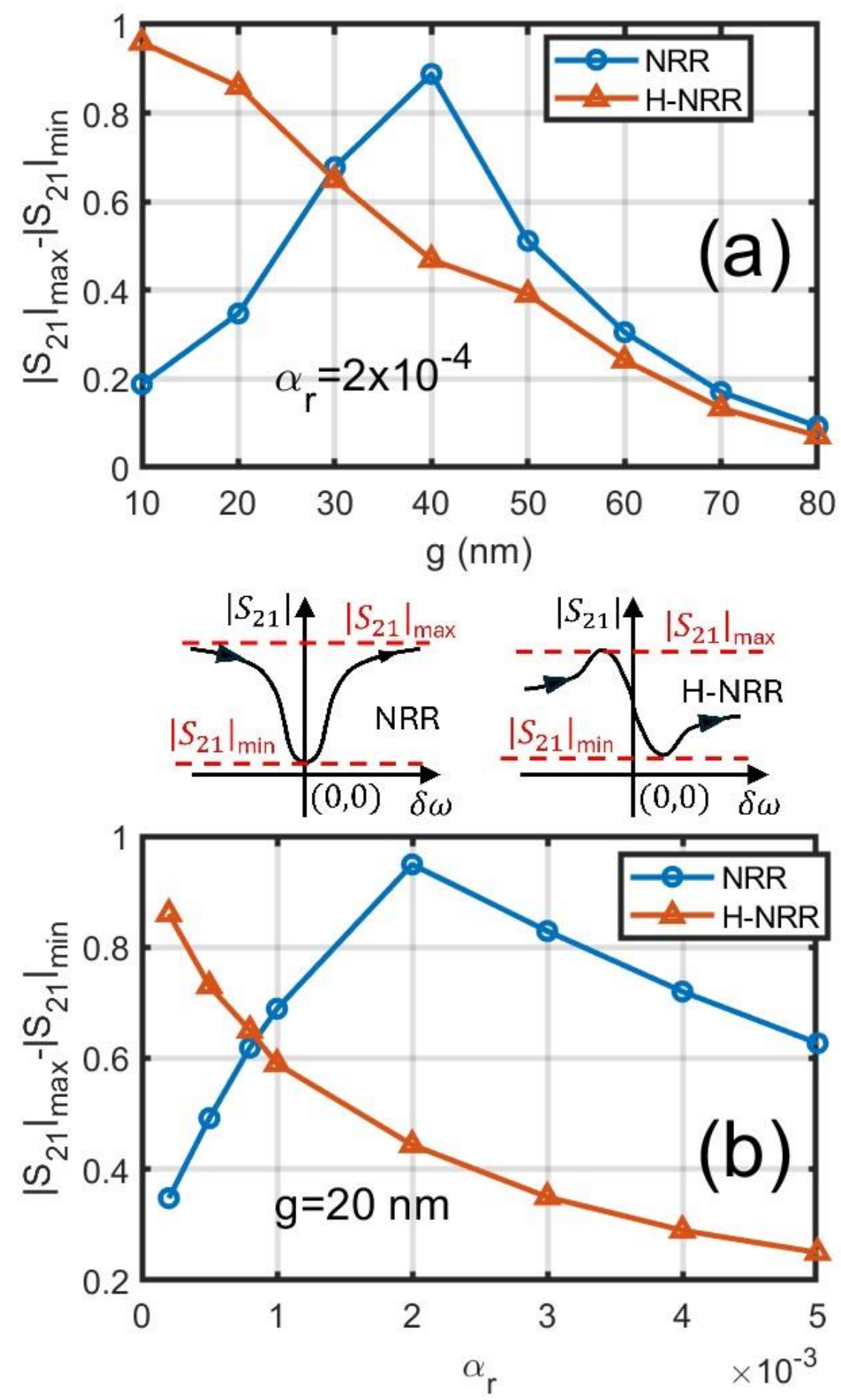

Figure 4 (a) Simulated results of transmission variation ($|S_{21}|_{max}$-$|S_{21}|_{min}$) vs. gap width (g) for a fixed $\alpha_r = 2 \times 10^{-4}$ for the case of NRR (blue circles) and H-NRR (brown triangles) where the hole size $d_h$=40 nm for the H-NRR. (b) Same as (a) but with gap g fixed at 20 nm while varying $\alpha_r$.

In conclusion an analytical model has been developed to describe the transmission and reflection characteristics of both NRR and H-NRR in the linear regime by considering the forward and backward propagation of spin-waves, and is validated by comprehensive micromagnetic simulation results. For magnonic application where a large transmission variation ($|S_{21}|_{max}$-$|S_{21}|_{min}$) against incidence spin-wave frequency is desired, our results show that H-NRR provides an alternative engineering process. Instead of in the case of NRR having to optimize parameters such as g or $\alpha_r$ for maximum transmission variation within an "increase-then-decrease" pattern, the H-NRR transmission variation has a monotonic dependence on these parameters, which provide more certainty and tolerance for experimentalists in fabrication.

**Supplementary material**

The supplementary material details the micromagnetic simulation methodology for the H-NRR and NRR using MuMax3. It examines the effects of shape, edge roughness, and hole displacement on the transmission and reflection spectra, along with energy conservation in the H-NRR. The magnetization distribution of the H-NRR is further supported by a quasi-static external-field study of the hysteresis loop of the coupled ring–waveguide system. In addition, the analytical model is supported by presenting the solution and outline of derivation.

**Acknowledgement**

This work is supported by funding from the Icelandic Research Fund Grant No. 228951 and 2410333.

**Disclosures.** The authors declare no conflicts of interest.

---

[1] Q. Wang, A. Hamadeh, R. Verba, V. Lomakin, M. Mohseni, B. Hillebrands, A. V Chumak, and P. Pirro, "A nonlinear magnonic nano-ring resonator," NPJ Comput. Mater. 6(1), 192 (2020).
[2] Y.-Y. Au, "Electric chiral magnonic resonators utilizing spin–orbit torques," J. Appl. Phys. **135**(4),043904 (2024).
[3] Y.-Y. Au, and K.G. Fripp, "Electric field control of chiral magnonic resonators for spin-wave manipulation," Phys. Rev. Appl. **20**(3), 034023 (2023).
[4] K.G. Fripp, Y. Au, A. V Shytov, and V. V Kruglyak, "Nonlinear chiral magnonic resonators: Toward magnonic neurons," Appl. Phys. Lett. 122(17), (2023).
[5] A. Lutsenko, K.G. Fripp, L. Flajšman, A. V Shytov, V. V Kruglyak, and S. van Dijken, "Magnonic Fabry–Pérot resonators as programmable phase shifters," Appl. Phys. Lett. **126**(8), (2025).
[6] H. Qin, R.B. Holländer, L. Flajšman, F. Hermann, R. Dreyer, G. Woltersdorf, and S. van Dijken, "Nanoscale magnonic Fabry-Pérot resonator for low-loss spin-wave manipulation," Nat. Commun. **12**(1), 2293 (2021).
[7] O.A. Santos, and B.J. van Wees, "Magnon confinement in an all-on-chip YIG cavity resonator using hybrid YIG/Py magnon barriers," Nano Lett. 23(20), 9303–9309 (2023).
[8] K.G. Fripp, A. V Shytov, and V. V Kruglyak, "Spin-wave control using dark modes in chiral magnonic resonators," Phys. Rev. B **104**(5), 054437 (2021).
[9] F. Vilsmeier, T. Taniguchi, M. Lindner, C. Riedel, and C. Back, "Spin wave propagation in a ring-shaped magnonic waveguide," Appl. Phys. Lett. **127**(16), (2025).
[10] M. Balynsky, Y. Khivintsev, A. Kozhevnikov, Y. Nikulin, V. Sakharov, Y. Filimonov, and A. Khitun, "Magnonic active ring co-processor," J. Appl. Phys. **133**(2), (2023).
[11] S.A. Odintsov, S.E. Sheshukova, S.A. Nikitov, and A. V Sadovnikov, "Lateral and vertical spin-wave transport in a multimode magnonic ring coupler," Phys. Rev. Appl. **22**(1), 014042 (2024).
[12] A.B. Ustinov, R. V Haponchyk, and M. Kostylev, "A current-controlled magnonic reservoir for physical reservoir computing," Appl. Phys. Lett. 124(4), (2024).
[13] D. Kazakov, T.P. Letsou, M. Beiser, Y. Zhi, N. Opačak, M. Piccardo, B. Schwarz, and F. Capasso, "Active mid-infrared ring resonators," Nat. Commun. 15(1), 607 (2024).
[14] N.L. Kazanskiy, S.N. Khonina, and M.A. Butt, "A review of photonic sensors based on ring resonator structures: three widely used platforms and implications of sensing applications," Micromachines (Basel). 14(5), 1080 (2023).

[15] V. Fallahi, Z. Kordrostami, and M. Hosseini, “Sensitivity and quality factor improvement of photonic crystal sensors by geometrical optimization of waveguides and micro-ring resonators combination,” Sci. Rep. 14(1), 2001 (2024).

[16] D.G. Rabus, C. Sada, D.G. Rabus, and C. Sada, “Ring resonators: Theory and modeling,” Integrated Ring Resonators: A Compendium, 3–46 (2020).

[17] M.F. Limonov, M. V Rybin, A.N. Poddubny, and Y.S. Kivshar, “Fano resonances in photonics,” Nat. Photonics 11(9), 543–554 (2017).

[18] M.F. Limonov, “Fano resonance for applications,” Adv. Opt. Photonics **13**(3), 703–771 (2021).

[19] A.E. Miroshnichenko, S. Flach, and Y.S. Kivshar, “Fano resonances in nanoscale structures,” Rev. Mod. Phys. 82(3), 2257–2298 (2010).

[20] L. Gu, H. Fang, J. Li, L. Fang, S.J. Chua, J. Zhao, and X. Gan, “A compact structure for realizing Lorentzian, Fano, and electromagnetically induced transparency resonance lineshapes in a microring resonator,” Nanophotonics **8**(5), 841–848 (2019).

[21] Z. Tu, D. Gao, M. Zhang, and D. Zhang, “High-sensitivity complex refractive index sensing based on Fano resonance in the subwavelength grating waveguide micro-ring resonator,” Opt. Express 25(17), 20911–20922 (2017).

[22] L. Gu, L. Fang, H. Fang, J. Li, J. Zheng, J. Zhao, Q. Zhao, and X. Gan, “Fano resonance lineshapes in a waveguide-microring structure enabled by an air-hole,” APL Photonics **5**(1), (2020).

[23] J. Wang, J. Lin, P. Jin, S. Liu, and K. Zhou, “Fano resonance in a microring resonator with a micro-reflective unit,” Opt. Express 31(19), 31587–31596 (2023).

[24] A. Vansteenkiste, J. Leliaert, M. Dvornik, M. Helsen, F. Garcia-Sanchez, and B. Van Waeyenberge, “The design and verification of MuMax3,” AIP Adv., vol. 4, no. 10, 2014.

[25] T. S. Safi, P. Zhang, Y. Fan, Z. Guo, J. Han, E. R. Rosenberg, C. Ross, Y. Tserkovnyak, and L. Liu, “Variable spincharge conversion across metal-insulator transition”, Nat. Comm.. **11**, 1 (2020)

[26] A. A. Nikitin and E. Lähderanta, “Spin-wave signal rejection in magnonic waveguides induced by conducting films with a single groove,” Phys. Rev. Appl. **22**, 064049 (2024)

[27] A. Krysztofik, N. Kuznetsov, H. Qin, L. Flajšman, E. Coy, and S. van Dijken, “Tuning of magnetic damping in Y3Fe5O12/metal bilayers for spin-wave conduit termination,” Materials **15**(8), 2814 (2022)

**Supplementary Material for "Fano Resonance of a Magnonic Ring Resonator with a Nano-Scatterer Embedded in a Magnonic Conduit"**

Rajesh Kumar[1,*], Yat-Yin Au[1], and Snorri Ingvarsson[1]

[1] Science Institute, University of Iceland, Dunhaga 3, IS-107 Reykjavik, Iceland

*rak43@hi.is

## A Size and Shape of Hole in Waveguide of H-NRR

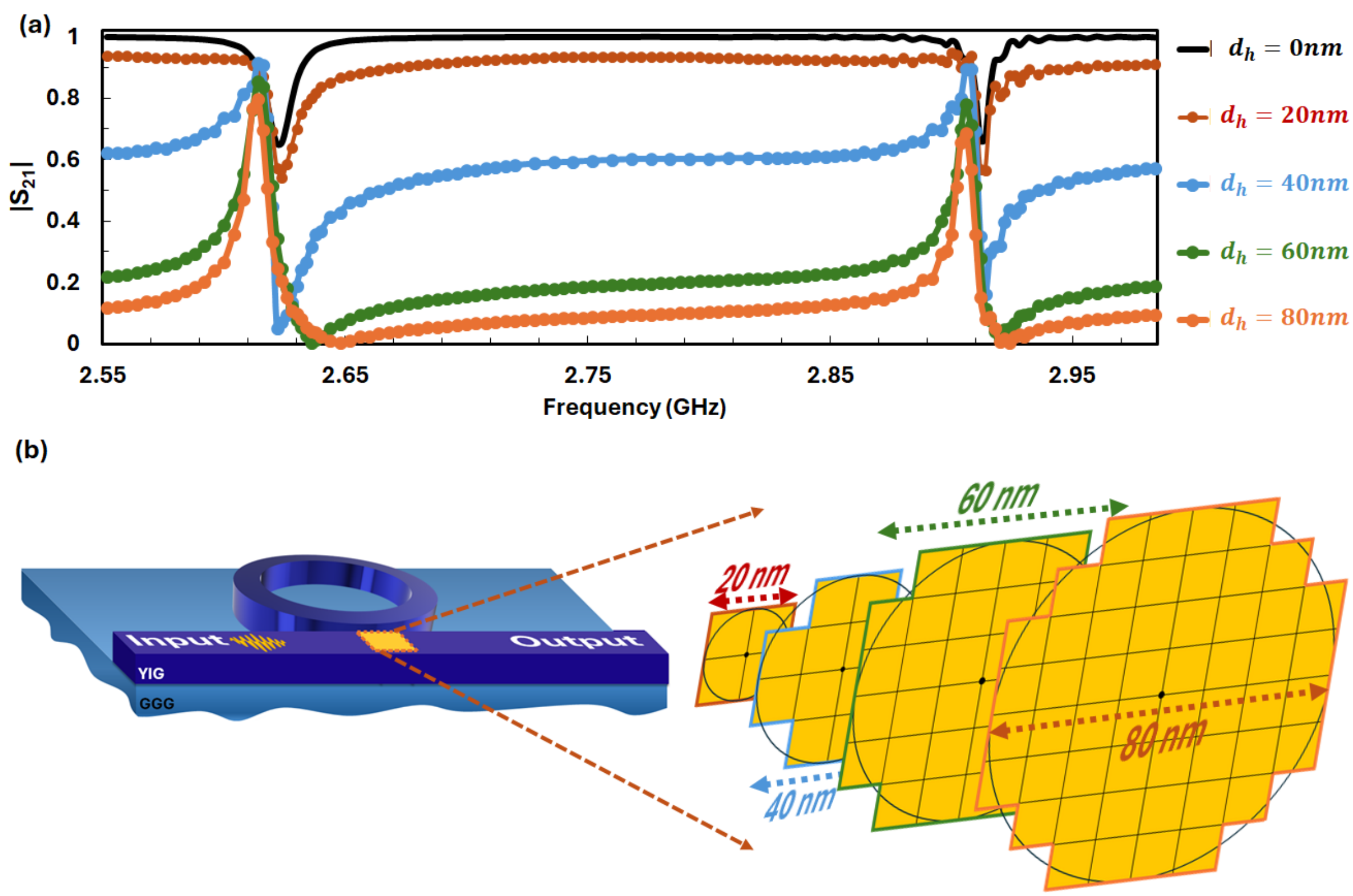

Figure 1 (a) Transmission spectra of the H-NRR for different $d_h$. (b) Grid-discretized circular boundaries of the hole for different $d_h$ (20nm, 40nm, 60nm and 80nm).

**Figure 1 Alt Text: Figure 1 has two panels (a to b)**

Panel (a) plots transmission $|S_{21}|$ (0 to 1) versus frequency (2.55 to 2.98 GHz) for hole sizes $d_h = 0,20,40,60,$ and 80 nm, showing a transition from symmetric Lorentzian dips to asymmetric Fano line shapes and lower off-resonance transmission as $d_h$ increases. Panel (b) illustrates a 3D schematic of the nanoring on a YIG/GGG waveguide with an enlarged inset displaying four grid-discretized circular meshes representing the hole geometries for 20 nm, 40 nm, 60 nm, and 80 nm.

To analyse the effect of hole size in the H-NRR, the transmission spectra were investigated by varying the hole grid-discretized diameter $d_h$ from 0 nm to 80 nm in steps of 20 nm (Figure *1*a). The results show that as the hole diameter increases, the transmission of spin waves at the off-resonance frequencies gradually decreases. As the hole diameter increases from 0 nm to 40 nm, the resonance lineshape transitions from a Lorentzian profile to a Fano profile at the resonance frequency; with further increase, the Fano lineshape becomes slightly distorted.

In MuMax3, the shape of a circular hole depends on the discretization determined by the computational cell size. The circular boundary is represented using discrete grid cells, where the occupancy of each cell determines whether it belongs to the hole or to the surrounding magnetic material. If the occupancy of a cell is less than 0.5, it is truncated to zero (not included), whereas if it is greater than 0.5, it is rounded up to unity (included). In our simulations, the cell size is $10 \times 10 \times 10$ nm$^3$. For a hole with $d_h = 20$ nm, all boundary cells have an occupancy greater than 0.5, resulting in an effective square-shaped hole with a side length of 20 nm. In contrast, for $d_h = 40$ nm, some boundary cells have occupancies less than 0.5 while others exceed 0.5, leading to a semi-square–like boundary approximation (Figure *1*b). Figure *1*b also illustrates the grid-discretized circular boundaries for $d_h$ of 20 nm, 40 nm, 60 nm, and 80 nm. To obtain a well-defined Fano lineshape for further investigation, $d_h = 40$nm was selected (shown in the main manuscript).

## B Micromagnetic Simulation

To investigate the spin-wave profiles, magnetization dynamics, and the transmission and reflection spectra of the NRR and H-NRR, micromagnetic simulations were performed using MuMax3, which solves the Landau–Lifshitz–Gilbert equation. A broadband pulse excitation (BPE) technique was employed to roughly determine the resonance frequencies in the transmission and reflection spectra, while a continuous wave excitation (CWE) technique was used to obtain the more accurate spectra and analyse the spin-wave profiles and magnetization dynamics at specific frequencies. The magnetization parameters of YIG are taken from experimental data[1], such as the exchange constant $A_{ex}$ = 3.5 pJm$^{-1}$ , saturation magnetization $M_s$= $1.4 \times 10^5$ Am$^{-1}$ and the Gilbert damping factor of the waveguides is $\alpha_{wg} = 2 \times 10^{-4}$.

The simulation uses a cell size of $10 \times 10 \times 10$ nm$^3$, which is smaller than the exchange length of YIG ($L = \sqrt{A_{ex}/\mu_0 M_s^2} \approx 16$nm), ensuring accurate resolution of the magnetization dynamics.  The static magnetization of the waveguide is uniformly aligned along its length (x-axis) due to strong shape anisotropy arising from its small width and thickness relative to its length. In contrast, the magnetization in the ring is initialized in a vortex state with chirality $-1$(clockwise) and polarity $+1$ (the polarity is irrelevant in this case since it is a ring and the ring centre is empty). Such a vortex configuration can be readily achieved by appropriately varying the applied magnetic field (see Section C). Subsequently, the system is allowed to relax to reach its ground state in the absence of an external magnetic field. To calculate the frequency spectra and magnetization dynamics, the micromagnetic simulations are divided into three parts: reference simulation, perfect reflection simulations and scattering simulations. In the reference simulations (ref) and perfect reflection simulation (pft), a reference waveguide and a perfectly reflecting waveguide are modelled, as shown in Table 1 Parameters used in the micromagnetic simulations a and b, respectively. In the scattering simulations (sct), a waveguide with a resonator is considered (Table 1 Parameters used in the micromagnetic simulations c), where the structures could either be NRR or H-NRR. The spatial distribution of the Gilbert damping factor ($\alpha$) along the $x$-direction is identical for both the ref and sct case. Specifically, $\alpha = \alpha_{wg}$ in the region between $x_3$ and $x_4$, while at both ends of the waveguide, $\alpha$ is gradually increased from $\alpha_{wg}$to $\alpha_d$ (where, $\alpha_d = \alpha_{wg} + 2\alpha_{dmp}$ , with $\alpha_{dmp} = 0.1$) to suppress spin-wave reflections (Table 1 Parameters used in the micromagnetic simulations d). In the pft simulation, the distribution of $\alpha$ along $x$ is the same as that of the ref simulation in the left half of the waveguide. However, in the right half (from $x_0$to $x_4$), $\alpha$ is set to a very high value, $\alpha_h = 10^6$ (Table 1 Parameters used in the micromagnetic simulations d), ensuring that the spin wave is fully reflected at position $x_0$ . The positions $x_{tdr}$ , $x_1$ , and $x_2$ correspond to the transducer, input, and output regions, respectively. The numerical values of all $x$-positions are provided in Table 1. To generate spin waves in the continuous wave excitation (CWE) mode, a uniform sinusoidal magnetic field $b_{rf} = b_0 \sin(2\pi f t)$is applied over the transducer region (green cuboid in Table 1 Parameters used in the micromagnetic simulations a to c), with the microwave frequency $f$ varied and the amplitude set to $b_0 = 0.1$ mT. In the broadband pulse excitation (BPE) mode, a sinc-shaped pulse is used as the excitation field, with its spatial and temporal profile, $b_z(x, t)$, detailed in Table 1. To calculate the transmission and reflection, the magnetization along the z-direction averaged over the cross section of the waveguide ($\overline{m}_z$) is recorded after the system has attained steady state. The specific $t_{max}$values for each mode are also provided in Table 1. At position $x_1$ (as input) and $x_2$ (as output), $\overline{m}_z$ for each simulation obtained as $\overline{m}_{z,sct}(x_1, t)$, $\overline{m}_{z,sct}(x_2, t)$, $\overline{m}_{z,ref}(x_1, t)$, $\overline{m}_{z,ref}(x_2, t)$ , $\overline{m}_{z,pft}(x_1, t)$ , and $\overline{m}_{z,pft}(x_2, t)$. In the BPE, transmission ($S_{21}$) and reflection ($S_{11}$) is calculated as

$$S_{21} = \frac{\mathcal{F}(\overline{m}_{z,sct}(x_2,t))(f)}{\mathcal{F}(\overline{m}_{z,ref}(x_2,t))(f)} \quad (1)$$

$$S_{11} = -\frac{\mathcal{F}(\overline{m}_{z,sct}(x_1,t) - \overline{m}_{z,ref}(x_1,t))(f)}{\mathcal{F}(\overline{m}_{z,pft}(x_1,t) - \overline{m}_{z,ref}(x_1,t))(f)} \quad (2)$$

Where $\mathcal{F}$ denotes the Fourier transforms from time domain to frequency domain. In the CWE, magnitude and phase of transmission and reflection are defined as:

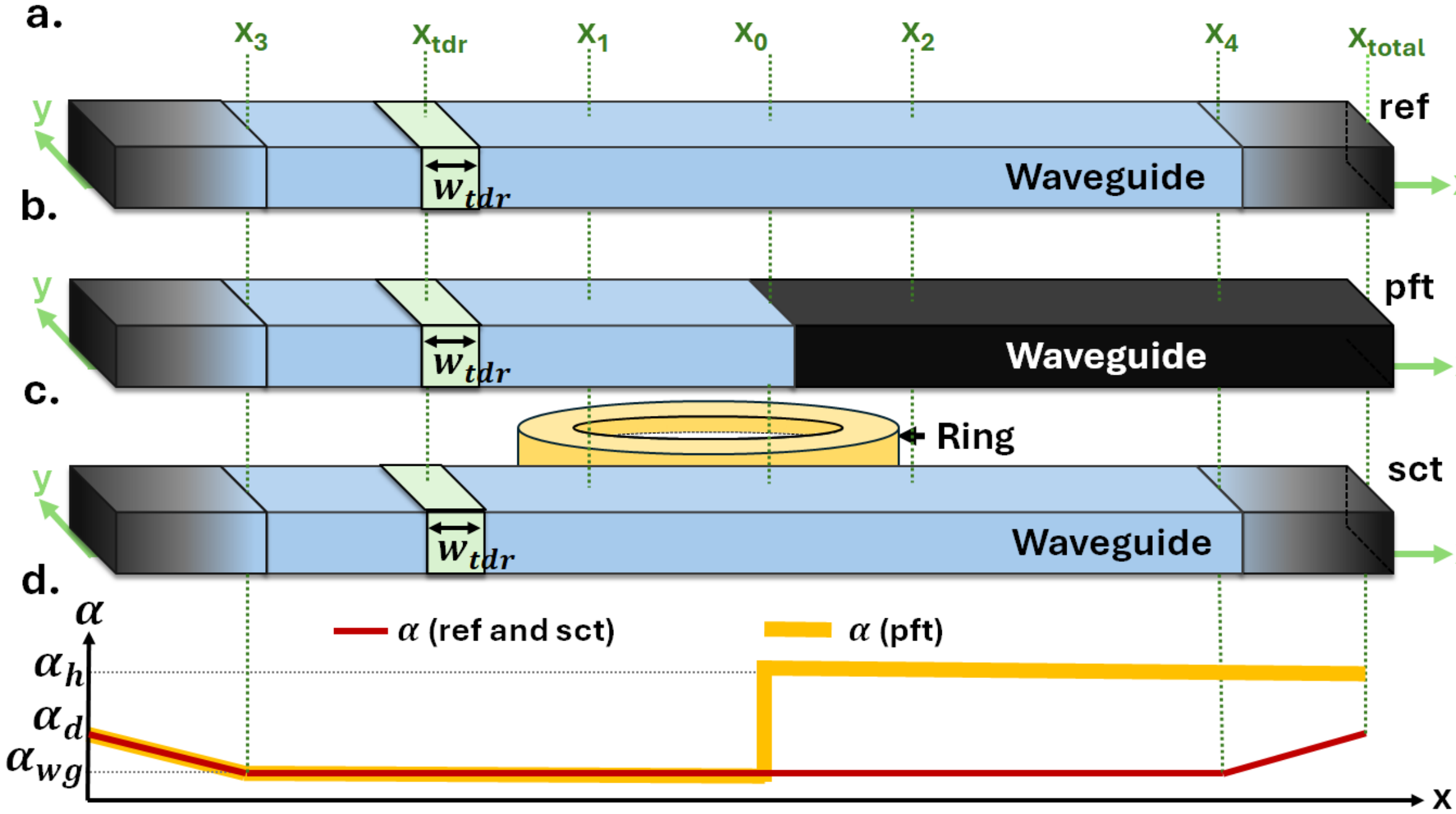

*Figure 2* Three-dimensional schematic of the micromagnetic models: (a) reference waveguide (ref), (b) perfectly reflecting waveguide (pft), and (c) waveguide with ring resonator (sct). (d) Spatial distribution of the Gilbert damping factor $\alpha$ as a function of $x$. This configuration is applicable to both BPE and CWE simulations.

**Figure 2 Alt Text: Figure 2 has 4 panels (a to d)**

Panels (a)–(c) illustrate 3D schematics of three waveguide models aligned along spatial positions $x_3$, $x_{\text{tdr}}$, $x_1$, $x_0$, $x_2$, $x_4$, and $x_{\text{total}}$: a reference waveguide (ref) with a transducer region of width $w_{\text{tdr}}$ (a), a perfectly reflecting waveguide (pft) with a dark absorber section starting at $x_0$ (b), and a scattering waveguide (sct) coupled to a yellow ring resonator at $x_2$ (c). Panel (d) plots the spatial distribution of the Gilbert damping factor $\alpha$ along position $x$ to define the numerical boundary conditions. It contrasts the uniform low baseline damping $\alpha_{\text{wg}}$ of the reference and scattering models (red curve) against a sharp step up to high damping $\alpha_h$ starting at $x_0$ for the perfectly reflecting model (yellow curve). Gradual absorbing ramps up to $\alpha_d$ are incorporated at both outer edges ($x < x_3$ and $x > x_4$) to prevent unwanted wave reflections at the simulation boundaries.

$$|S_{21}| = \frac{\text{mag}\{\overline{m}_{z,sct}(x_2,t)\}}{\text{mag}\{\overline{m}_{z,ref}(x_2,t)\}} \quad (3)$$

$$|S_{11}| = \frac{\text{mag}(\overline{m}_{z,sct}(x_1,t) - \overline{m}_{z,ref}(x_1,t))}{\text{mag}(\overline{m}_{z,pft}(x_1,t) - \overline{m}_{z,ref}(x_1,t))} \quad (4)$$

$$\angle S_{21} = \text{phase}(\overline{m}_{z,sct}(x_2,t) - \overline{m}_{z,ref}(x_2,t)) \quad (5)$$

$$\angle S_{11} = \pi + \text{phase}\{\overline{m}_{z,sct}(x_1,t) - \overline{m}_{z,ref}(x_1,t)\} - \text{phase}\{\overline{m}_{z,pft}(x_1,t) - \overline{m}_{z,ref}(x_1,t)\} \quad (6)$$

Here, $|S_{21}|$ and $|S_{11}|$ represent the magnitudes of transmission and reflection, while $\angle S_{21}$ and $\angle S_{11}$ denote the phases of transmission and reflection, respectively. mag{} and phase{} correspond to the resultant amplitude and phase obtained from the sinusoidal fit of $\overline{m}_z$ once the system reaches a steady state.

| **Parameter** | **BPE** | **CWE** |
|---|---|---|
| $x_{total}$ | 8 μm | 4000 nm |
| $x_0$ | 4 μm | 2000 nm |
| $x_1$ | 3.1 μm | 1600 nm |
| $x_2$ | 3.9 μm | 2400 nm |
| $x_3$ | 2.0 μm | 800 nm |
| $x_4$ | 6.0 μm | 3200 nm |
| $x_{tdr}$ | 2.5 μm | 1500 nm |
| $b_z(x,t)$ | $b_0 \frac{\sin(\omega_c(t-t_0))}{\omega_c(t-t_0)} \frac{\sin(k_c(x-x_{tdr}))}{k_c(x-x_{tdr})} - \frac{\omega_{c2}}{\omega_c} b_0 \frac{\sin(\omega_{c2}(t-t_0))}{\omega_{c2}(t-t_0)} \frac{\sin(k_{c2}(x-x_{tdr}))}{k_{c2}(x-x_{tdr})}$ | $h_0 \sin(\omega t)\cos(2\pi(x - x_{tdr})/w_{tdr})\min(1.0, 1.1 \times (1-e^{-t/t_{rise}}))$ |
| $w_{tdr}$ | 1 μm | $\lambda_{cw}$ |
| $t_{max}$ | 100 ns | 150 ns |
| $t_{rise}$ | 4 ns | |
| $t_0$ | 1 ns | |
| $f_c (= \omega_c/(2\pi))$ | 4 GHz | |
| $\lambda_c = (2\pi/k_c)$ | 140 nm | |
| $f_{c2} (= \omega_{c2}/(2\pi))$ | 2.5 GHz | |
| $\lambda_{c2} = (2\pi/k_{c2})$ | 500 nm | |

Table 1 Parameters used in the micromagnetic simulations.

**Table 1 Alt Text:**

A two-column comparison table presents micromagnetic simulation parameters for Broadband Pulse Excitation (BPE) and Continuous Wave Excitation (CWE) methods. The spatial domain parameters for BPE versus CWE are total length $x_{\text{total}}$ ($8\ \mu$m vs. $4000$ nm), absorber origin $x_0$ ($4\ \mu$m vs. $2000$ nm), coupling start position $x_1$ ($3.1\ \mu$m vs. $1600$ nm), resonator center $x_2$ ($3.9\ \mu$m vs. $2400$ nm), left damping ramp end $x_3$ ($2.0\ \mu$m vs. $800$ nm), right damping ramp start $x_4$ ($6.0\ \mu$m vs. $3200$ nm), and transducer center $x_{\text{tdr}}$ ($2.5\ \mu$m vs. $1500$ nm). For the excitation field $b_z(x,t)$, BPE uses a dual sinc-function pulse in space and time with transducer width $w_{\text{tdr}} = 1\ \mu$m, total time $t_{\text{max}} = 100$ ns, rise time $t_{\text{rise}} = 4$ ns, pulse center $t_0 = 1$ ns, upper cutoff frequency $f_c = 4$ GHz ($\lambda_c = 140$ nm), and lower cutoff frequency $f_{c2} = 2.5$ GHz ($\lambda_{c2} = 500$ nm). In contrast, CWE utilizes a sinusoidal continuous wave excitation formula modulated by a spatial cosine function and exponential rise term, with transducer width defined by $\lambda_{\text{cw}}$ and a total simulation duration $t_{\text{max}} = 150$ ns.

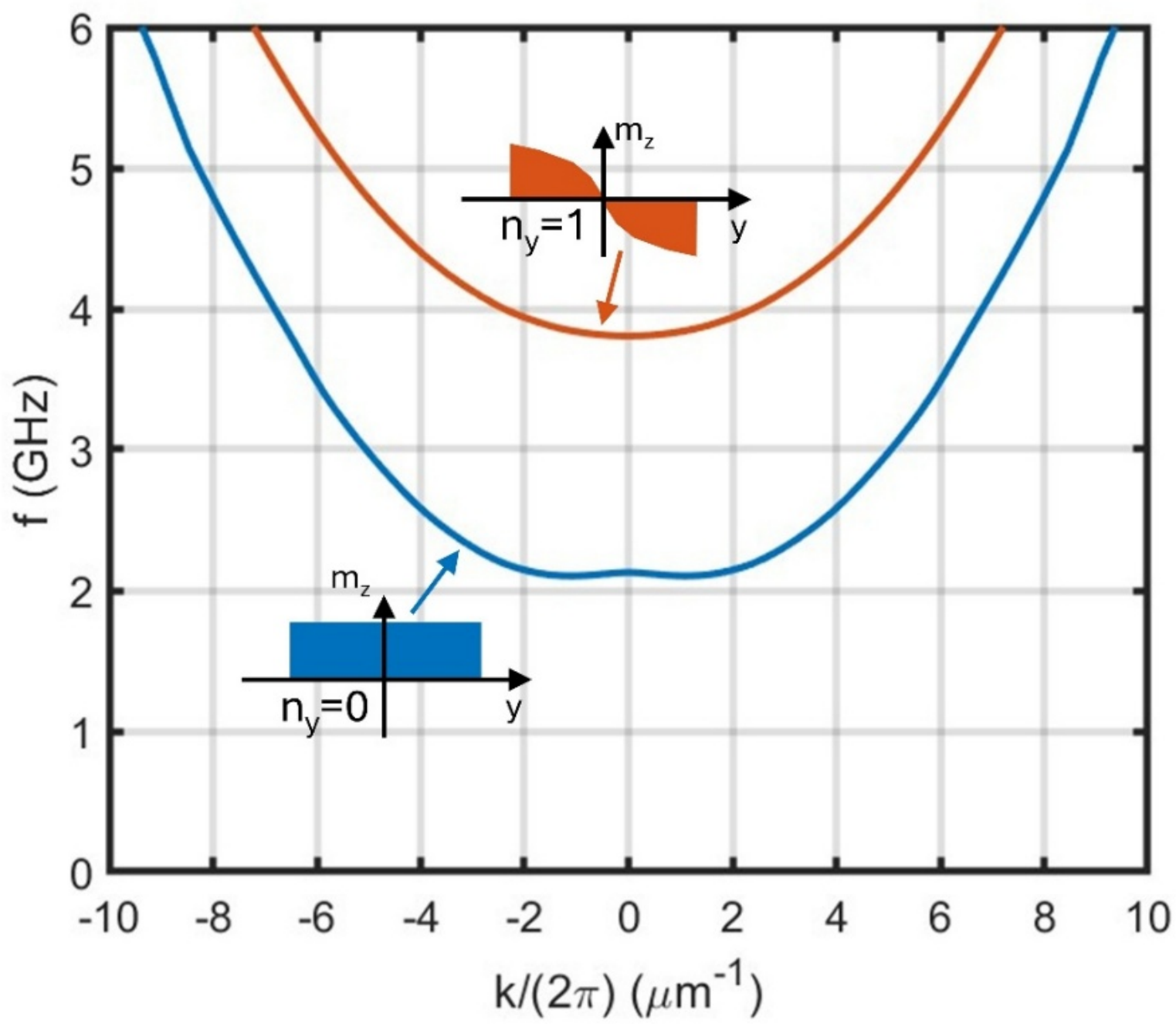

*Figure 3* Dispersion relation of spin wave mode inside a 100 nm wide and 30 nm thick YIG waveguide conduit.

**Figure 3 Alt Text:**

A line plot displays the spin-wave dispersion relation plotting frequency $f$ in GHz (0 to 6) against wavenumber $k/(2\pi)$ in $\mu$m$^{-1}$ ($-10$ to 10) for a 100 nm wide, 30 nm thick YIG waveguide. The plot features two parabolic, symmetric curves corresponding to the first two transverse spin-wave modes: a lower blue curve for the $n_y = 0$ mode with a frequency minimum near 2.1 GHz at $k = 0$, and an upper brown curve for the $n_y = 1$ mode with a higher band-edge frequency minimum near 3.8 GHz at $k = 0$. Both dispersion curves exhibit a parabolic shape that increases monotonically in frequency as the absolute value of the wavenumber $|k|$ increases toward $\pm 10\ \mu$m$^{-1}$.

To calculate the $\mathrm{n_y}$ th branch dispersion relation in the YIG waveguide, an excitation field given by

$$\mathrm{b_z(x,t) = b_0 \frac{\sin(\omega_c(t-t_0))}{\omega_c(t-t_0)} \frac{\sin(k_c(x-x_0))}{k_c(x-x_0)} sgn\{\cos[\frac{2\pi}{\lambda_y} y]\}} \tag{7}$$

is excited in a ref simulation (which contains only the waveguide), where $\lambda_y = 2w_{wg}/n_y$. Dispersion relation (f vs k) is then obtained from the maxima of $|M_z(f,k)|$, where

$$M_z(f,k) = \mathcal{F}\{\mathcal{F}\{m_z(x, y_{line}, z_{line}, t)\}(f)\}(k) \quad (8)$$

In equation (8), $m_z$ is the resultant z component magnetic moment dynamics of the YIG waveguide as a function of position and time. Also,

$$y_{line} = \begin{cases} y_{total}/2 & \text{for } n_y \text{ is even} \\ y_{total}/2 - \lambda_y/4 & \text{for } n_y \text{ is odd} \end{cases} \quad (9)$$

and $z_{line} = d/2$. Also, $\mathcal{F}\{\}(f)$ and $\mathcal{F}\{\}(k)$ denote Fourier transform from t to f and from $x$ to $k$. The above simulation leads to dispersion relationship results displayed in Figure 3 for a $d = 30$ nm thick and $w_{wg} = 100$ nm wide YIG conduit for the $n_y$=0 and $n_y$=1 mode branch. One can observe from Figure 3 that spin wave in the waveguide remain single mode for f < 3.8 GHz.

## C Quasi-static external field study of the ring plus waveguide system hysteresis loop

To explain the preparation of the ground state for subsequent dynamic study mentioned in Section B, we study the hysteresis loop behaviour of the ring plus waveguide system under quasi static cycling external field applied in x direction. In Figure 4a, the applied magnetic field started at $B_{ext,x}$=+50 mT and was slowly decreased in value. During this process relaxation of magnetic moment from the saturation state happens (Figure 4b) due to shape anistropy inside the ring. When $B_{ext,x}$ is further decreased to -18 mT, a domain wall is formed inside the ring in the region next to the waveguide (Figure 4c). This domain wall is eliminated when $B_{ext,x}$ is further decreased to -23 mT (Figure 4d). This vortex state is short lived and is destroyed after $B_{ext,x}$ is decreased beyond -25 mT, where the ring transitions into a largely homogeneous state (with moments point towards negative x). The huge transition of $M_x$ from +0.55 to -1 at $B_{ext,x}$=-40 mT corresponds to the flipping of magnetic moment inside the waveguide (magnetization state not shown).

In order to prepare the ring to be in a vortex state with magnetic moments aligned anti-parallel with those inside the waveguide, as required in the dynamic studies in the main article, we take advantage of the hysteresis loop presented in Figure 4a. After the transition from domain wall state to vortex state, instead of further decreasing $B_{ext,x}$, the field is ramped up back to zero (Figure 5a), which result at the target state of magnetization stated in the beginning of this paragraph (Figure 5e).

To achieve a vortex state in the ring with magnetic moment parallel to the waveguide, one should again start from $B_{ext,x}$=+50 mT, but decrease the field up to a point that magnetization is completely reversed and then ramp back upward (Figure 6a and b). Then use the same trick as in Figure 5 to achieve a vortex state (Figure 6d) and slowly decrease the $B_{ext,x}$ to zero. The target magnetization stated in the beginning of this paragraph is thus achieved (Figure 6e).

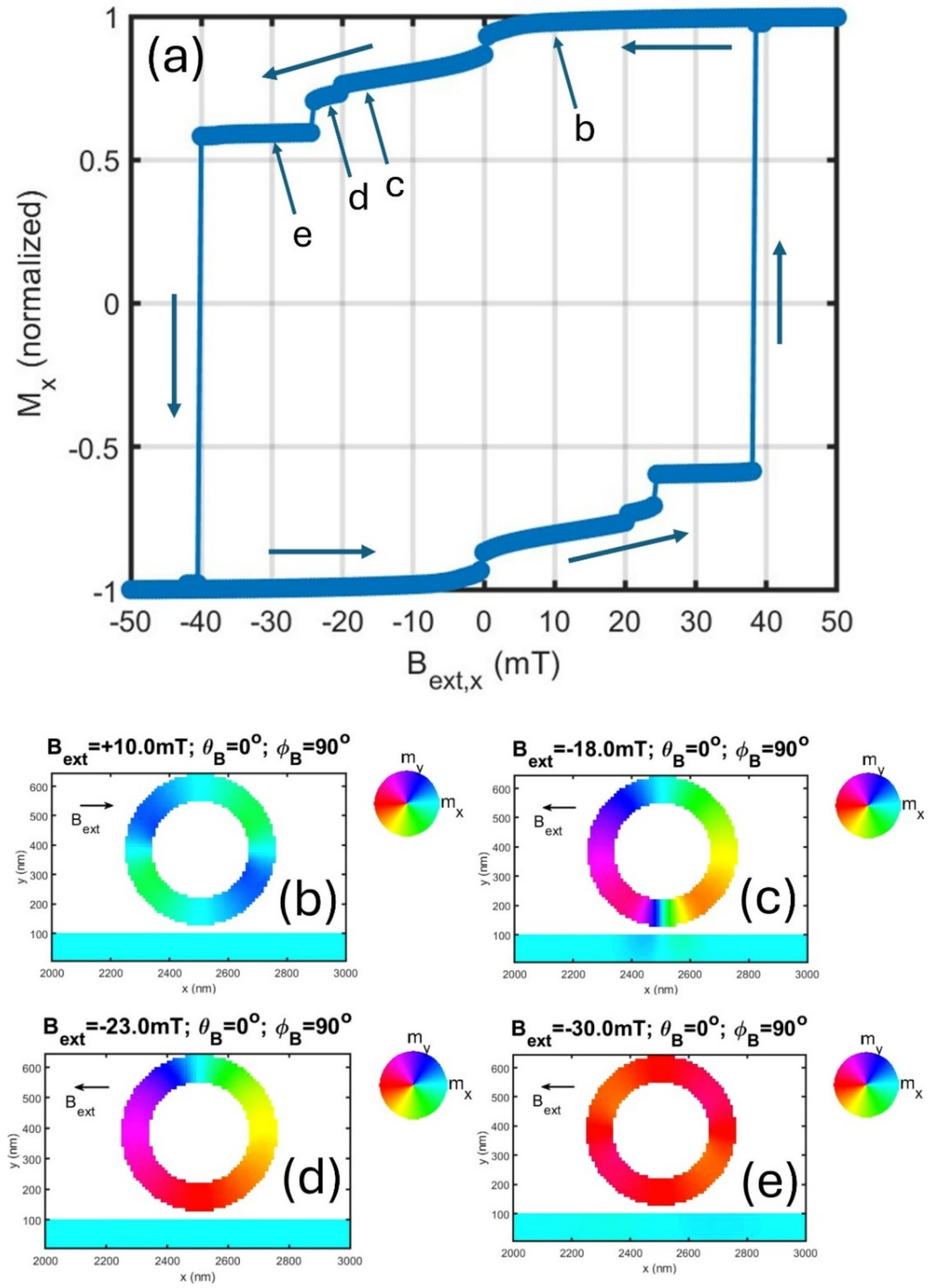

*Figure 4* (a) Hysteresis loop of normalized sum of all x component magnetic moment in the ring and the waveguide ($M_x$) against applied external magnetic field in x direction ($B_{ext,x}$). The state of magnetization at certain position on the hysteresis loop are displayed in (b) to (e).

**Figure 4 Alt Text: Figure 4 has five panels (a to e)**

Panel (a) displays a magnetic hysteresis loop plotting normalized magnetization $M_x$ (vertical axis, $-1$ to $1$) against external magnetic field $B_{\text{ext},x}$ in mT (horizontal axis, $-50$ to $50$ mT) for the ring and waveguide system, showing distinct stepped transitions marked with labels b, c, d, and e. Panels (b) through (e) show 2D in-plane magnetization maps using a rainbow color wheel corresponding to specific field points along the decreasing field branch: panel (b) at $B_{\text{ext}} = +10.0$ mT shows an asymmetric onion-like magnetization state with $M_x \approx 0.95$; panel (c) at $B_{\text{ext}} = -18.0$ mT illustrate change in magnetization direction at coupling region( $M_x \approx 0.72$ ); panel (d) at $B_{\text{ext}} = -23.0$ mT exhibits further switching to an intermediate state ($M_x \approx 0.65$); and panel (e) at $B_{\text{ext}} = -30.0$ mT shows a stable, reverse-oriented magnetization state ($M_x \approx 0.58$). Across all field states, the adjacent waveguide maintains a uniform cyan color indicating fixed rightward magnetization.

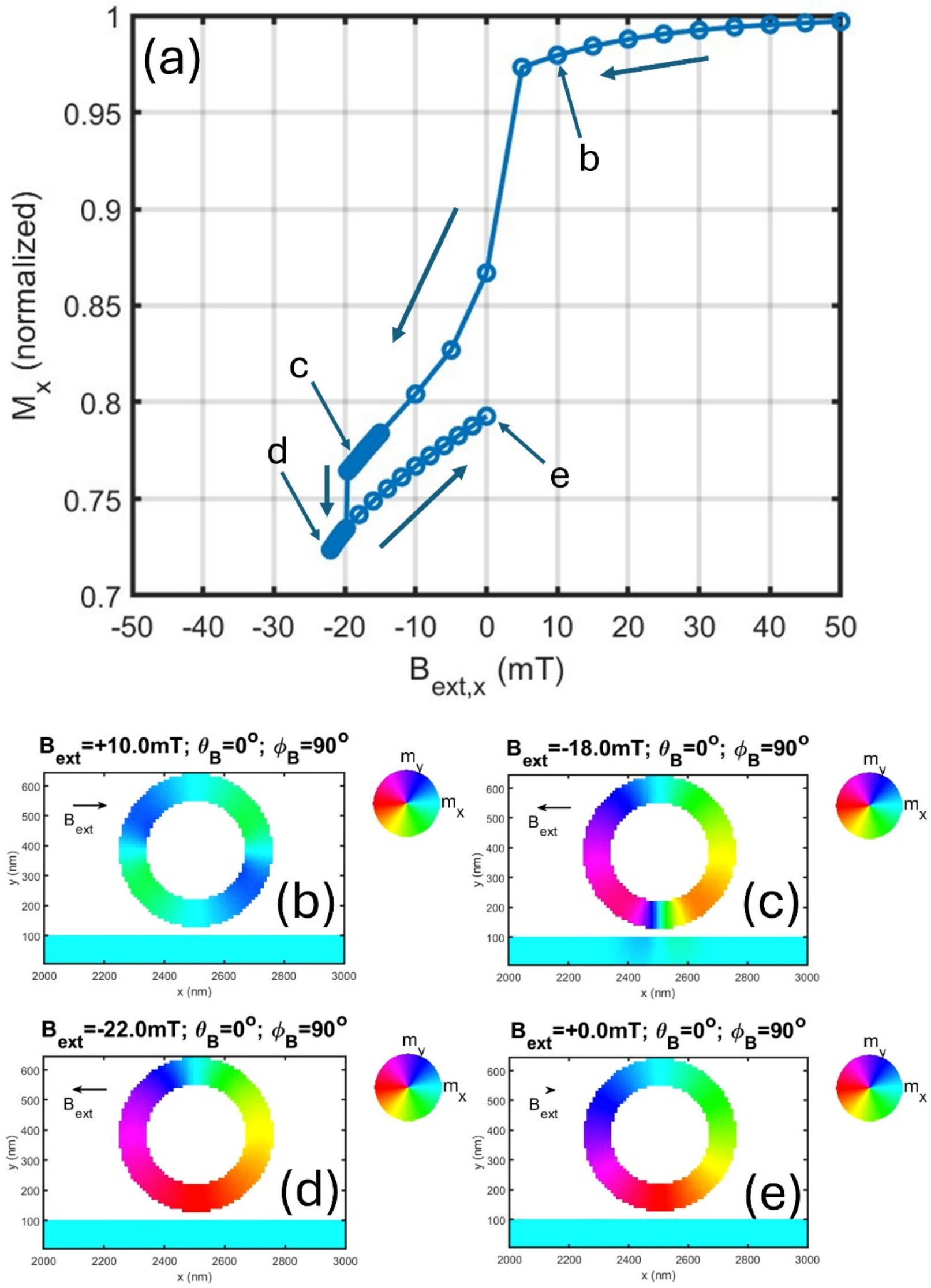

*Figure 5* (a) Normalized sum of all x component magnetic moment in the ring and the waveguide ($M_x$) against applied external magnetic field in x direction ($B_{ext,x}$). The state of magnetization at certain position on the hysteresis loop are displayed in (b) to (e). The applied field procedure is aiming at achieving the ring to be in a vortex state with anti-parallel alignment of magnetic moment between the ring and the waveguide at the coupling region.

**Figure 5 Alt Text: Figure 5 has five panels (a to e)**

Panel (a) plots normalized magnetization $M_x$ (vertical axis, $0.7$ to $1.0$) against applied external field $B_{\text{ext},x}$ in mT (horizontal axis, $-50$ to $50$ mT), showing a minor magnetization loop designed to stabilize a vortex ground state. As $B_{\text{ext},x}$ decreases from $+50$ mT, $M_x$ drops from 1.0 down to $\sim 0.73$ at $B_{\text{ext}} = -22.0$ mT, before recovering along an upper minor branch back to $\sim 0.79$ at zero field ($B_{\text{ext}} = 0$ mT). Panels (b)–(e) illustrate 2D spatial magnetization maps corresponding to labeled points on the loop: panel (b) at $B_{\text{ext}} = +10.0$ mT displays an asymmetric onion state ($M_x \approx 0.98$); panel (c) at $B_{\text{ext}} = -18.0$ mT illustrate the change in magnetization direction near the ring edge ($M_x \approx 0.77$); panel (d) at $B_{\text{ext}} = -22.0$ mT displays a fully formed vortex state ($M_x \approx 0.73$); and panel (e) at $B_{\text{ext}} = 0.0$ mT demonstrates the remanent flux-closure vortex state ($M_x \approx 0.79$) with anti-parallel magnetic alignment relative to the cyan waveguide at the coupling interface.

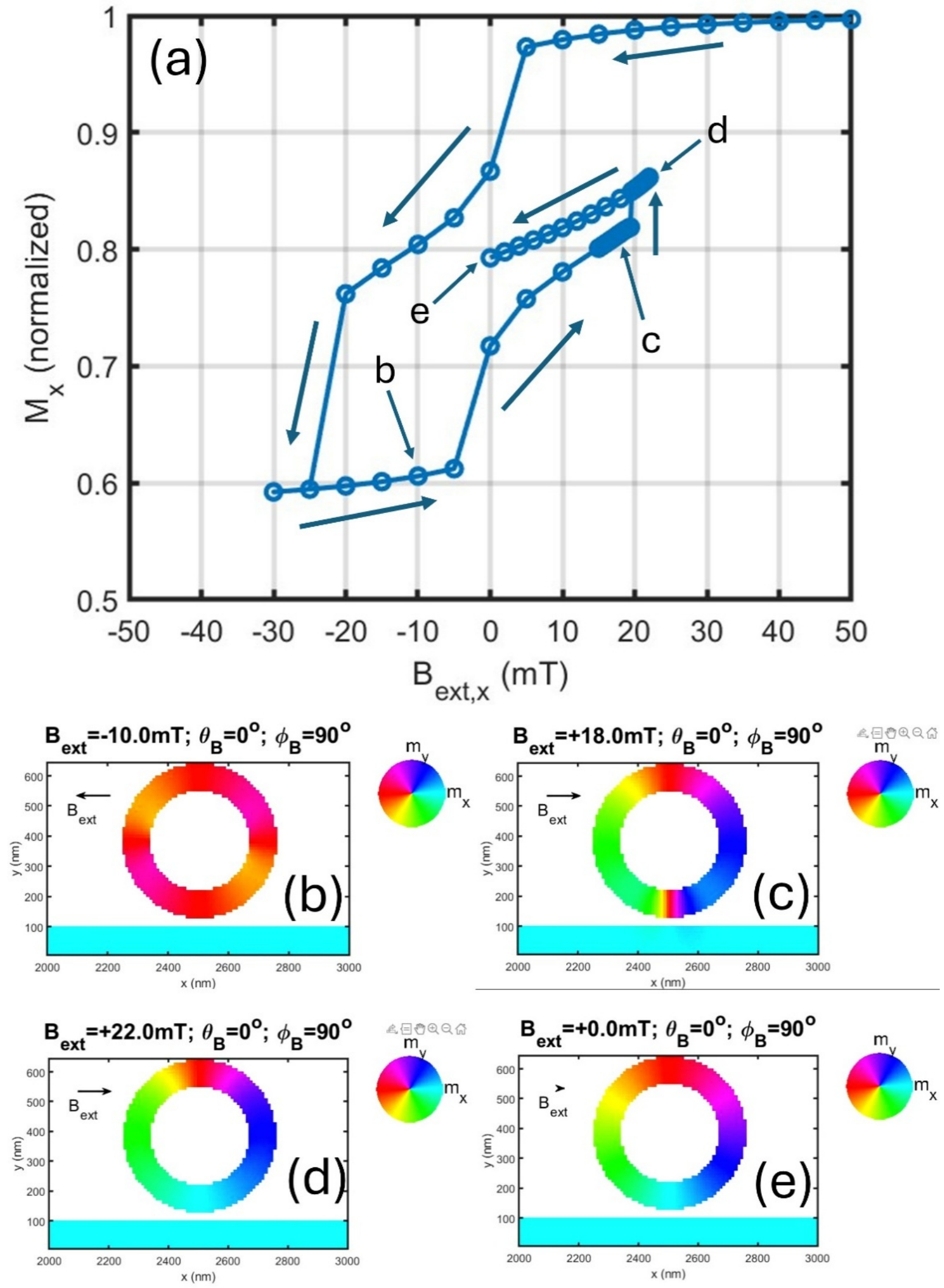

Figure 6 (a) Normalized sum of all x component magnetic moment in the ring and the waveguide ($M_x$) against applied external magnetic field in x direction ($B_{ext,x}$). The state of magnetization at certain position on the hysteresis loop are displayed in (b) to (e). The applied field procedure is aiming at achieving the ring to be in a vortex state with parallel alignment of magnetic moment between the ring and the waveguide at the coupling region.

**Figure 6 Alt Text: Figure 6 has five panels (a to e)**

Panel (a) plots normalized magnetization $M_x$ (vertical axis, $0.5$ to $1.0$) against applied external field $B_{\text{ext},x}$ in mT (horizontal axis, $-50$ to 50 mT), depicting a minor magnetization loop designed to achieve a parallel-aligned vortex state. As $B_{\text{ext},x}$ decreases from $+50$ mT into negative fields, $M_x$ drops to a minimum baseline of $\sim 0.59$ near $-30$ mT, then increases as the field is ramped back toward positive values, reaching $\sim 0.81$ at $B_{\text{ext}} = +18.0$ mT (c) and $\sim 0.86$ at $+22.0$ mT (d), before settling on an upper minor branch. Panels (b)–(e) demonstrate 2D spatial magnetization maps using a rainbow color wheel corresponding to labeled points on the loop: panel (b) at $B_{\text{ext}} = -10.0$ mT displays a reverse-oriented onion state ($M_x \approx 0.60$); panel (c) at $B_{\text{ext}} = +18.0$ mT illustrate change in magnetization direction near the coupling gap ($M_x \approx 0.81$); panel (d) at $B_{\text{ext}} = +22.0$ mT displays a fully formed vortex state ($M_x \approx 0.86$); and panel (e) at $B_{\text{ext}} = 0.0$ mT illustrates the remanent flux-closure vortex state ($M_x \approx 0.79$) with parallel magnetic alignment relative to the cyan waveguide at the coupling interface.

## D Solution and outline of its derivation for the theory

The solution to the two tracks model in the main article can be given as $r = r_s + (\psi_L/\psi_0)$ and $\tau = \tau_s + (\psi_R/\psi_0)$, where

$$\psi_R = [c_R e^{-ik\eta/2} + \tau_s c_R + r_s c_L](-e^{-ik\eta/2} + 1) \tag{10}$$

$$\psi_L = [c_L e^{-ik\eta/2} + \tau_s c_L + r_s c_R](-e^{-ik\eta/2} + 1) \tag{11}$$

$$c_R = -\frac{2}{\eta k} e^{ik\eta/2} \frac{\psi_0}{A^2 - B^2} [(\Gamma_F + \Gamma_B)\{A\alpha_1 - B\alpha_2\} + 2\sqrt{\Gamma_F\Gamma_B}\{A\alpha_2 - B\alpha_1\}] \tag{12}$$

$$c_L = -\frac{2}{\eta k} e^{ik\eta/2} \frac{\psi_0}{A^2 - B^2} [2\sqrt{\Gamma_F\Gamma_B}\{A\alpha_1 - B\alpha_2\} + (\Gamma_F + \Gamma_B)\{A\alpha_2 - B\alpha_1\}] \tag{13}$$

$$A = (\omega - \Omega_0 + i\Gamma_0) + (\Gamma_F + \Gamma_B)\beta + 2\sqrt{\Gamma_F\Gamma_B}\gamma \tag{14}$$

$$B = 2\sqrt{\Gamma_F\Gamma_B}\beta + (\Gamma_F + \Gamma_B)\gamma \tag{15}$$

$$\alpha_1 = \frac{1}{\eta}[\frac{1 - e^{-ik\eta/2}}{ik} + \tau_s \frac{e^{ik\eta/2} - 1}{ik}] \tag{16}$$

$$\alpha_2 = \frac{1}{\eta} r_s \frac{e^{ik\eta/2} - 1}{ik} \tag{17}$$

$$\beta = \frac{2e^{ik\eta/2}}{\eta^2 k}[-\eta e^{-ik\eta/2} + \frac{1 - e^{-ik\eta/2}}{ik} + \{\tau_s(-e^{-ik\eta/2} + 1) + e^{-ik\eta/2}\}\frac{e^{ik\eta/2} - 1}{ik}] \tag{18}$$

$$\gamma = \frac{2e^{ik\eta/2}}{\eta^2 k} r_s(-e^{-ik\eta/2} + 1)\frac{e^{ik\eta/2} - 1}{ik} \tag{19}$$

Equations (10) to (19) form multiple layers of encapsulation, which ultimately express $r$ and $\tau$ as a function of $\omega$ and $k$, with $\Omega_0$, $\Gamma_0$, $\Gamma_F$, $\Gamma_B$, $\eta$, $r_s$ and $\tau_s$ as fitting parameters. At the same time, $\omega$ and $k$ are related to each other through the dispersion relation given in Figure 3. The proof of the solution given by (10) to (19) are as followed: Suppose the full solution to equation (3) and (4) in the main article to be given as

$$\phi_{R(L)} = \phi_{0R(L)} e^{-i\omega t} \tag{20}$$

$$\psi_R(x,t) = \begin{cases} \psi_0 e^{ikx} e^{-i\omega t} & \text{for } x < -\eta/2 \\ [\psi_0 e^{ikx} + \psi_{R\Delta}(x)] e^{-i\omega t} & \text{for } -\eta/2 < x < 0 \\ [\tau_s \psi_0 e^{ikx} + \psi_{R\Delta}(x)] e^{-i\omega t} & \text{for } 0 < x < \eta/2 \\ [\tau_s \psi_0 + \psi_R] e^{ikx} e^{-i\omega t} & \text{for } x > \eta/2 \end{cases} \tag{21}$$

$$\psi_L(x,t) = \begin{cases} [r_s \psi_0 + \psi_L] e^{-ikx} e^{-i\omega t} & \text{for } x < -\eta/2 \\ [r_s \psi_0 e^{-ikx} + \psi_{L\Delta}(x)] e^{-i\omega t} & \text{for } -\eta/2 < x < 0 \\ \psi_{L\Delta}(x) e^{-i\omega t} & \text{for } 0 < x < \eta/2 \\ 0 & \text{for } x > \eta/2 \end{cases} \tag{22}$$

where

$$\psi_{R\Delta}(x) = \begin{cases} \dfrac{\eta_0(\Delta_F \phi_{0R} + \Delta_B \phi_{0L})}{v\eta k} + c_R e^{ikx} & \text{for } -\eta/2 < x < 0 \\ \dfrac{\eta_0(\Delta_F \phi_{0R} + \Delta_B \phi_{0L})}{v\eta k} + c_{Rs} e^{ikx} & \text{for } 0 < x < \eta/2 \end{cases} \tag{23}$$

$$\psi_{L\Delta}(x) = \begin{cases} \dfrac{\eta_0(\Delta_F \phi_{0L} + \Delta_B \phi_{0R})}{v\eta k} + c_{Ls} e^{-ikx} & \text{for } -\eta/2 < x < 0 \\ \dfrac{\eta_0(\Delta_F \phi_{0L} + \Delta_B \phi_{0R})}{v\eta k} + c_L e^{-ikx} & \text{for } 0 < x < \eta/2 \end{cases} \tag{24}$$

Applying continuity of $\psi_R(x)$ at $x = -\eta/2$, one obtains

$$c_R = -\frac{\eta_0(\Delta_F \phi_{0R} + \Delta_B \phi_{0L})}{v\eta k} e^{ik\eta/2} \tag{25}$$

Applying continuity of $\psi_L(x)$ at $x = +\eta/2$, one obtains

$$c_L = -\frac{\eta_0(\Delta_F \phi_{0L} + \Delta_B \phi_{0R})}{v\eta k} e^{ik\eta/2} \tag{26}$$

On the other hand, applying the scattering condition at $x = 0$ (equation (6) in the main article)

$$\begin{cases} \psi_{R\Delta}(0^+) = \tau_s \psi_{R\Delta}(0^-) + r_s \psi_{L\Delta}(0^+) \\ \psi_{L\Delta}(0^-) = \tau_s \psi_{L\Delta}(0^+) + r_s \psi_{R\Delta}(0^-) \end{cases} \tag{27}$$

One obtains

$$c_{Rs} = \tau_s \left(-e^{-ik\eta/2} + 1\right) c_R + r_s \left(-e^{-ik\eta/2} + 1\right) c_L + c_R e^{-ik\eta/2} \tag{28}$$

and

$$c_{Ls} = \tau_s\left(-e^{-ik\eta/2} + 1\right)c_L + r_s\left(-e^{-ik\eta/2} + 1\right)c_R + c_L e^{-ik\eta/2} \quad (29)$$

Now applying the continuity of $\psi_R(x)$ at $x = +\eta/2$, and continuity of $\psi_L(x)$ at $x = -\eta/2$, one obtains equations (10) and (11). On the other hand, substituting equations (20), (21) and (22) into equation (4) of the main article for $\phi_R$, one finally obtains

$$\begin{aligned}\left[(\omega - \Omega_0 + i\Gamma_0) + (\Gamma_F + \Gamma_B)\beta + 2\sqrt{\Gamma_F\Gamma_B}\gamma\right]\phi_{0R} + \left[2\sqrt{\Gamma_F\Gamma_B}\beta + (\Gamma_F + \Gamma_B)\gamma\right]\phi_{0L} \\ = \psi_0\{\Delta_F^*\alpha_1 + \Delta_B^*\alpha_2\}\end{aligned} \quad (30)$$

In between, we have assumed that $\Delta_F^*\Delta_B$ is a real number, thus $\eta_0\Delta_F^*\Delta_B/(2v) = \sqrt{\Gamma_F\Gamma_B}$. On the other hand, substituting equations (20), (21) and (22) into equation (4) of the main article for $\phi_L$, one eventually obtains

$$\begin{aligned}\left[2\sqrt{\Gamma_F\Gamma_B}\beta + (\Gamma_F + \Gamma_B)\gamma\right]\phi_{0R} + \left[(\omega - \Omega_0 + i\Gamma_0) + (\Gamma_F + \Gamma_B)\beta + 2\sqrt{\Gamma_F\Gamma_B}\gamma\right]\phi_{0L} \\ = \psi_0\{\Delta_B^*\alpha_1 + \Delta_F^*\alpha_2\}\end{aligned} \quad (31)$$

where $\alpha_1$, $\alpha_2$, $\beta$ and $\gamma$ are given by equations (16) to (19). Solving equations (30) and (31) for $\phi_{0R}$ and $\phi_{0L}$, one obtains

$$\begin{bmatrix}\phi_{0R}\\ \phi_{0L}\end{bmatrix} = \frac{\psi_0}{A^2 - B^2}\begin{bmatrix}A & -B\\ -B & A\end{bmatrix}\begin{bmatrix}\{\Delta_F^*\alpha_1 + \Delta_B^*\alpha_2\}\\ \{\Delta_B^*\alpha_1 + \Delta_F^*\alpha_2\}\end{bmatrix} \quad (32)$$

where $A$ and $B$ are given by equation (14) and (15). Substituting equation (32) into (25) and (26), one obtains equations (12) and (13). Also, by comparing equations (21) and (22) with (7) and (8) in the main article, one finds $r = r_s + (\psi_L/\psi_0)$ and $\tau = \tau_s + (\psi_R/\psi_0)$.

In case $r_s = 0$ and $\tau_s = 1$ (i.e. no micro-scatterer), the explicit forms of $r$ and $\tau$ can be given as

$$r = \frac{-i4\{\omega - \Omega_0 + i\Gamma_0\}\sqrt{\Gamma_F\Gamma_B}\zeta''}{\{\omega - \Omega_0 + i\Gamma_0 - (\Gamma_F + \Gamma_B)\zeta\}^2 - (2\sqrt{\Gamma_F\Gamma_B}\zeta)^2} \quad (33)$$

$$\tau = 1 - i\left(\frac{\{\omega - \Omega_0 + i\Gamma_0\}2(\Gamma_F + \Gamma_B)\zeta'' - 2\zeta\zeta''(\Gamma_F - \Gamma_B)^2}{\{\omega - \Omega_0 + i\Gamma_0 - (\Gamma_F + \Gamma_B)\zeta\}^2 - \left(2\sqrt{\Gamma_F\Gamma_B}\zeta\right)^2}\right) \quad (34)$$

where

$$\zeta = \zeta' - i\zeta'' = \frac{2}{(k\eta)^2}\{k\eta - \sin(k\eta)\} - i\frac{\sin^2(k\eta/2)}{(k\eta/2)^2} \tag{35}$$

Therefore, if the parameter $k\eta$ is large enough, $r$ and $\tau$ can exhibit a double peak and double dip feature, which is a manifestation of self-hybridization of the two degenerated modes in the resonator. On the other hand, it is worth mentioning that in case where is only one mode ($\phi$) in the resonator instead of two ($\phi_R$ and $\phi_L$), the coefficients $r$ and $\tau$ are reduced to the following form

$$r = \frac{-i2\sqrt{\Gamma_F\Gamma_B}\zeta''}{\omega - \Omega_0 + i\Gamma_0 - (\Gamma_F + \Gamma_B)\zeta} \tag{36}$$

$$\tau = 1 - \frac{i2\Gamma_F\zeta''}{\omega - \Omega_0 + i\Gamma_0 - (\Gamma_F + \Gamma_B)\zeta} \tag{37}$$

Therefore, the total linewidth of the Lorentzian in equations (36) and (37) is given as $\Gamma_{tot} = \Gamma_0 + (\Gamma_F + \Gamma_B)\zeta''$, thus $\Gamma'_{F(B)} = \zeta''\Gamma_{F(B)}$ could be seen as the effective radiative linewidth from the resonator into the waveguide broadened by the parameter $k\eta$. If $k\eta \to 0$, then $\zeta' \to 0$ and $\zeta'' \to 1$ and equations (36) and (37) reduce to their familiar form in Ref. 2.

## E Effect of position of hole in waveguide of H-NRR:

To study the effect of hole position, the transmission spectra of the H-NRR (with $d_h = 20nm$ and $\alpha_r = 1 \times 10^{-3}$) were analyzed by shifting the hole from the center toward the right side of the waveguide by a distance d (Figure 7a). The displacement of the air hole results in different asymmetric line shapes at the same resonant frequency. Figure 7b show the transmission spectra, as the hole is shifted from the center of the waveguide ($d = 0nm$) to $d = 200\,nm$, the Fano line shape gradually transforms into a Lorentzian line shape. As the hole moves away from the center of the waveguide, it eventually shifts out of the coupling region. As a result, the discrete resonant mode of the NRR couples back to the undisturbed continuum state of the waveguide and maintains a phase difference of 2π. When this interfering mode propagates beyond the coupling region, even though a hole is still present in the waveguide, both interfering components experience the same phase shift. Consequently, no additional phase difference is introduced, leading to the recovery of a Lorentzian resonance line shape. But due to the presence of the hole there is reduction in the transmission at the off-resonance frequency.

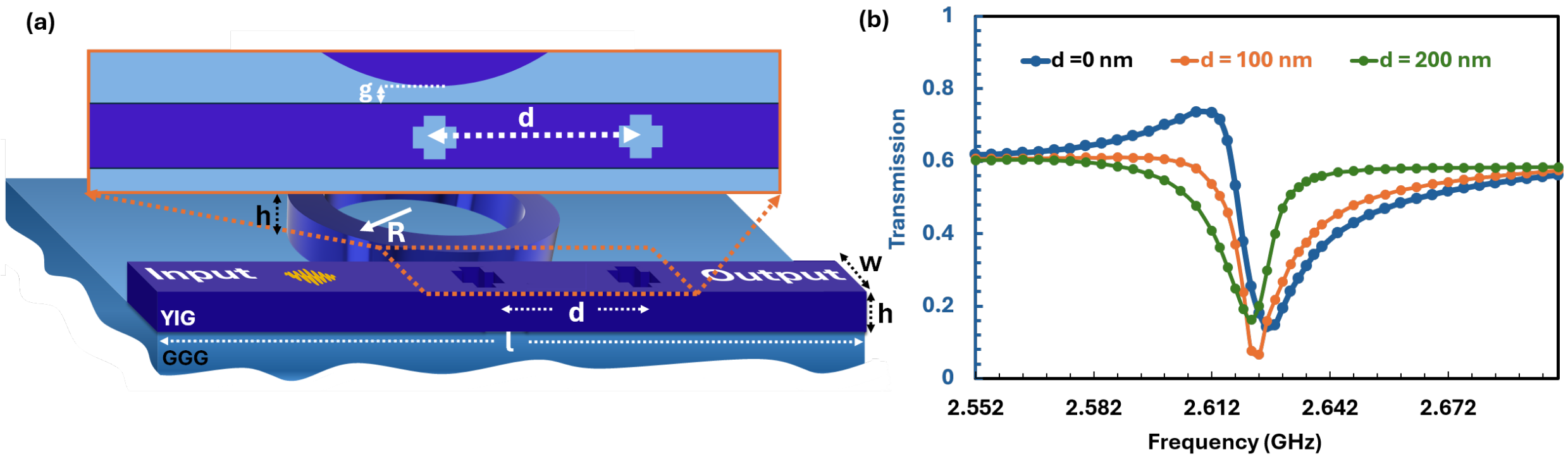

Figure 7 a) H-NRR showing shifting of the hole towards the right side. b) Transmission spectra for different d value.

Figure 7 Alt Text: Figure 7 has two panels (a to b)
Panel (a) displays a three-dimensional schematic of an H-NRR system placed in the xy-plane with its height oriented along the z-axis, featuring two cross-shaped micro-holes embedded in a YIG/GGG waveguide separated by a horizontal distance $d$, alongside structural dimensions including waveguide width $w$, height $h$, length $l$, ring radius $R$, and coupling gap $g$. Panel (b) plots transmission spectra showing transmission magnitude (vertical axis, 0 to 1) versus frequency in GHz (horizontal axis, 2.552 to 2.692 GHz) for three separation values: $d = 0$ nm (blue circles), $d = 100$ nm (orange circles), and $d = 200$ nm (green circles). As the separation distance $d$ increases, the spectrum shifts from an asymmetric Fano profile with a peak and deep dip ($d = 0$ nm) toward a more symmetric Lorentzian dip profile near 2.622 GHz for $d = 200$ nm.

## F Energy Conservation for the NRR and H-NRR

To investigate energy conservation, a simulation was performed with a ring diameter of 520 nm and $\alpha_r = 1 \times 10^{-3}$ for both the NRR and H-NRR structures. By extracting the magnitude of the complex amplitude ($|S_{21}|$ and $|S_{11}|$), energy conservation ($|S_{21}|^2 + |S_{11}|^2 < 1$) can be investigated. Figure 8a and b illustrate the spectra for transmission ($|S_{21}|$), reflection ($|S_{11}|$) and $|S_{21}|^2 + |S_{11}|^2$The $|S_{21}|^2 + |S_{11}|^2$ curve remains close to 1 at off-resonance frequencies, but drops sharply at the resonance frequency, indicating that energy is absorbed within the ring.

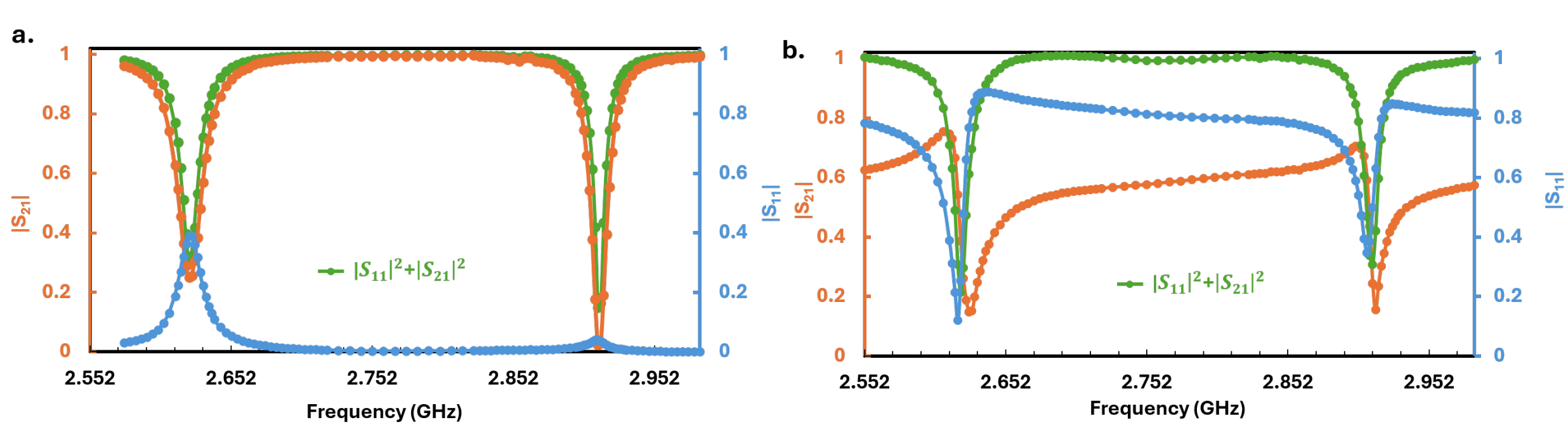

Figure 8 Simulated Transmission magnitude ($S_{21}$), reflection magnitude ($S_{11}$) spectra and plot of calculated $|S_{21}|^2 + |S_{11}|^2$ for: a) NRR and b) H-NRR.

**Figure 8 Alt Text: Figure 8 has two panels (a to b)**

Panels (a) and (b) plot the total calculated power sum $|S_{21}|^2 + |S_{11}|^2$ (green curve) against frequency (2.552 to 2.982 GHz) for NRR and H-NRR systems. In panel (a) for the standard NRR, the power sum maintains a flat baseline near 1.0 off-resonance with two sharp, symmetric dips down to $\sim 0.2$ at 2.62 GHz and 2.914 GHz. In panel (b) for the H-NRR, the power sum similarly stays near 1.0 off-resonance but exhibits broader, asymmetric dips at the same resonant frequencies as in panel (a).

## G Effect of hole shape randomness

In order to investigate the effect of shape randomness of the hole in the waveguide due to lithography imperfectness, we have simulated holes with various shapes as depicted in Figure 9(a) and the resultant $|S_{21}|$ spectra are displayed in Figure 9b. It is evidenced that as far as the randomness is concerned, it has very little impact on the transmission spectrum and the Fano lineshape. It is believed that the randomness defined by our lateral cell size (10×10 nm$^2$) is too small compared to the spin wave wavelength (~ 200 nm) to create a significant impact.

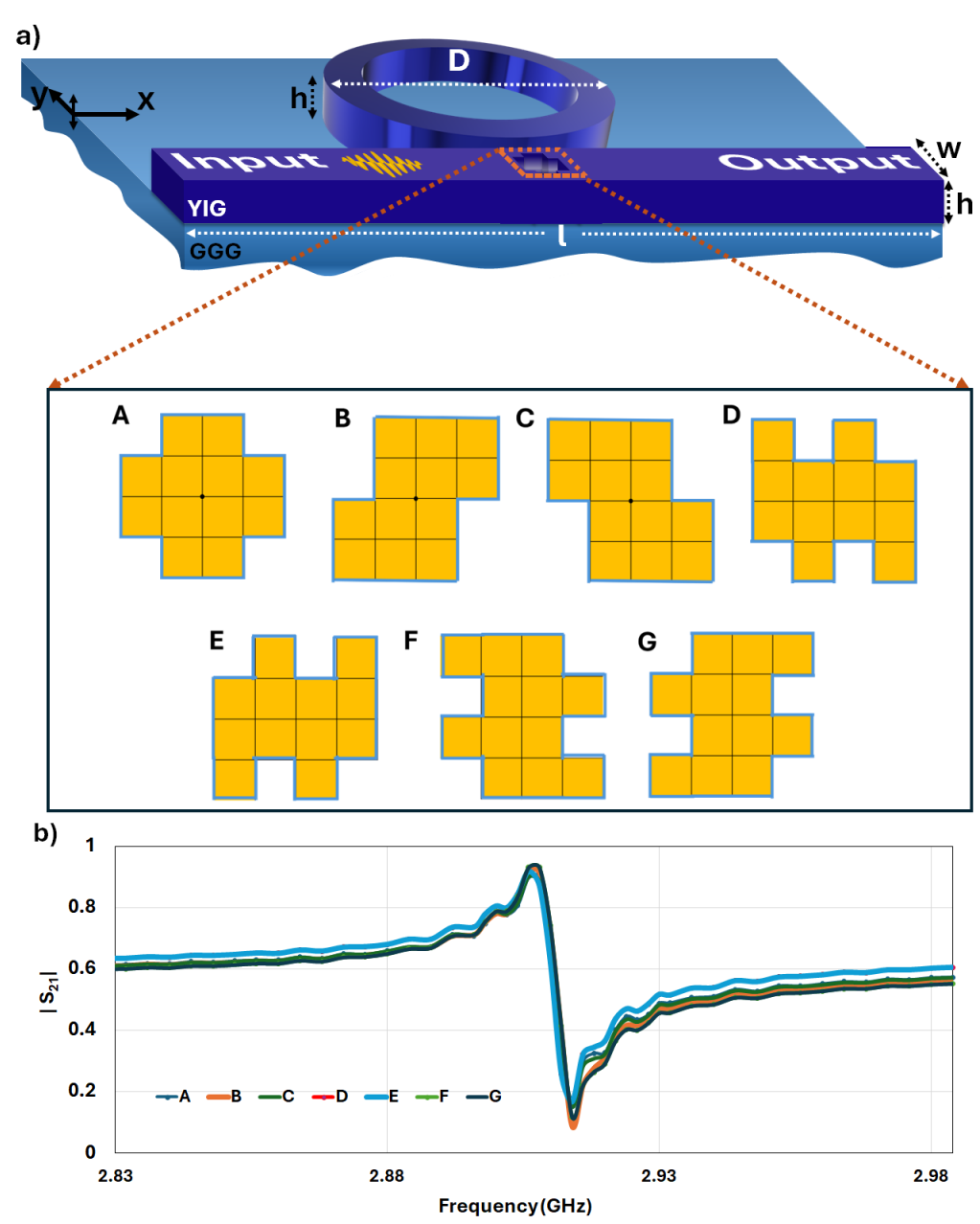

Figure 9 (a) Illustration of shape randomness of the hole embedded in the waveguide. Each square unit in A to G represent a 10×10 nm2 of void region in the micromagnetic simulation. (b) Resultant $|S_{21}|$ spectra for various shape of the hole.

**Figure 9 Alt Text: Figure 9 has two panels (a to b)**

Panel (a) presents a 3D schematic in the xy-plane illustrating a YIG/GGG waveguide (length $l$, width $w$, height $h$ along the z-axis) coupled to a ring resonator of outer diameter $D$, with an inset illustrating seven grid-based micro-hole geometries labeled A through G composed of $10 \times 10$ nm$^2$ void units to simulate shape randomness. Panel (b) plots the transmission spectra $|S_{21}|$ (vertical axis, 0 to 1) against frequency in GHz (horizontal axis, 2.83 to 2.98 GHz) for shapes A–G. All seven curves follow a nearly identical asymmetric Fano line shape with a transmission peak near 0.92 at 2.91 GHz followed by a sharp dip down to $\sim 0.1$ around 2.915 GHz, demonstrating robust spectral stability despite structural variations in hole shape.

[1] C. Dubs, O. Surzhenko, R. Thomas, J. Osten, T. Schneider, K. Lenz, J. Grenzer, R. Hübner, and E. Wendler, "Low damping and microstructural perfection of sub-40nm-thin yttrium iron garnet films grown by liquid phase epitaxy," Phys. Rev. Mater. 4(2), 024416 (2020).

[2] K.G. Fripp, A. V Shytov, and V. V Kruglyak, "Spin-wave control using dark modes in chiral magnonic resonators," Phys. Rev. B **104**(5), 054437 (2021).